\documentclass[a4paper,10pt]{article}
\usepackage{graphicx}
\usepackage{amsmath,amssymb,amsfonts,amsthm}
\usepackage{url, enumerate,anysize}
\usepackage{lscape}
\usepackage[subnum]{cases}
\usepackage{multirow}

\usepackage[usenames,dvipsnames,svgnames,table]{xcolor}
\usepackage{color}
\usepackage[colorlinks=true,linkcolor=black, citecolor=blue, urlcolor=blue]{hyperref}
\usepackage{graphicx}
\usepackage{tikz}
\usepackage[titletoc]{appendix}
\newtheorem{thm}{Theorem}[section]

\newtheorem{lem}[thm]{Lemma}

\usepackage[usenames,dvipsnames,svgnames,table]{xcolor}

\newcommand{\StatSpc}{\mathcal N}
\newcommand{\noi}{\noindent}

\newcommand{\prvV}{{\ddot V}}
\newcommand{\prvHO}{{\ddot H}_O }

\title{Speed Based Optimal Power Control in Small Cell Networks}

\author{Veeraruna Kavitha$^{1}$\footnote{Emails: V. Kavitha (vkavitha@iitb.ac.in), Manu K. Gupta (manu-kumar.gupta@irit.fr)}, Manu K. Gupta$^2$, V\'eronique Capdevielle$^{3}$, Rahul Kishor M$^{1}$ and Majed Haddad$^4$ \\ $^1$Industrial Engineering and Operations Research, IIT Bombay;\\ $^2$IRIT, 2 rue C. Camichel, Toulouse, France;\\$^3$Nokia, France; $^4$University of Avignon, France.}

\marginsize{1.5cm}{1.5cm}{1.5cm}{1.5cm}
\begin{document}
\maketitle 

\begin{abstract}
Small cell networks promise good quality of service (QoS) even for cell edge users, however pose challenges to cater to the high-speed users. The major difficulty being that of frequent handovers and the corresponding handover losses, which significantly depend upon the speed of the user. It was shown previously that the optimal cell size increases with speed. Thus, in scenarios with diverse users (speeds spanning over large ranges), it would be inefficient to serve all users using common cell radius and it is practically infeasible to design different cell sizes for different speeds. Alternatively, we propose to allocate power to a user based on its speed, e.g., higher power virtually increases the cell size.
We solve well known Hamiltonian Jacobi equations under certain assumptions to obtain a power law, optimal for load factor and busy probability, for any given average power constraint and cell size. The optimal power control turns out to be linear in speed.
We build a system level simulator for small cell network, using elaborate Monte-Carlo simulations, and show that the performance of the system improves significantly with linear power law.
The power law is tested even for the cases, for which the system does not satisfy the assumptions required by the theory.  
For example, the linear power law has significant improvement  in comparison with the 'equal power' system,  even in presence of time varying and random interference. 
We observe good improvement in almost all cases with improvements up to 89\% for certain configurations.
\end{abstract}


\section{Introduction}

Since the advent of smartphones, mobile networks are witnessing an exponential traffic growth. Cisco has recently predicted 11-fold increase in the global mobile data traffic between 2015 and 2020 (\cite{Cisco15}), while Qualcomm has forecasted an astounding 1000x increase in mobile data traffic in near future (\cite{Qualcomm13}). The increased traffic creates new challenges for service provider in various aspects, for example, mobility management (\cite{Perez12}).
To cater to the increasing traffic demands, cellular networks are transformed into heterogeneous networks (HetNets), by deploying additional small cells (pico cells, femto cells, metro cells, etc.) along with the existing macro cells (\cite{Nokia,Andrews}).
Sometimes, small base stations (BSs) are installed in high traffic sub-regions to share the load of the macro BS.
Major streets (high ways) are one such example sub regions, which often carry heavy traffic.
Proposals are made to install small BSs on the lamp posts along the side of major streets/high ways (e.g., \cite{Alcatel}).

Mobility management raises various novel research questions for short radius cells, more so when they cater to high speed users. As the cell size decreases, the handover rate critically increases especially for high or medium speed users which will increase the rate of call drops (\cite{STRAIGHT16}). Any handover is dropped if the entering new cell does not have a free server. Thus, the drop rate increases with the frequency of handovers.
On the other hand, with reduced cell radius, one can serve even the cell edge users with good service rate.


This trade-off (good service rates versus handover loss) is studied in \cite{Perf} and optimal cell size for appropriate performance measure (e.g., expected service time, load factor and drop or busy probability, etc.) is obtained. It was concluded that the optimal cell size for higher velocity users is larger. However, it would be a practically infeasible procedure to vary the physical cell size based on speed. 3GPP has proposed to offset the received signal power from various BSs to the user (under consideration) using a cell selection bias (CSB), before selecting the serving BS of the user. With CSB, one may not always select the BS with the strongest received power. The main idea is to favour the handover (HO) of high mobility users towards macro cells, while low or medium mobility users are offloaded to small cells. But this is a coarse (small or macro BS) control, which may not improve the performance significantly.

We propose to allocate power to the users based on their speed, which makes it possible to have a fine control.
The higher power virtually increases the cell size: a far away user with higher power can be served at the same transmission rate as a near-by user with an appropriate smaller power (capacity of the far away user increases with higher power) (\cite{Ref1}). This work is motivated by above natural phenomenon.
One can have as many levels of power allocation as required, enabling finer control.  We derive \textit{optimal speed-based power allocation} policies.

The user speed estimates have been recently studied by many researchers in small cell networks (\cite{Majed13,speed1, Majed16}). These estimates may (or may not) be available with high accuracy. If the estimates are not accurate, one can classify the users into finite number of classes (each class specified by a range of user speeds); allocate same power to all the users in a speed class. We derive a closed form expression for an optimal power control vector, whose components represent the power allocated to different speed classes, subject to a constraint on the average power used. We refer this as \textit{discrete} power law. Alternatively, if the user speed can be estimated accurately, we obtain the optimal power law using Hamilton-Jacobi Bellman (HJB) equations and Lagrange multipliers as function of user-speed. We refer this as \textit{continuous} power law.

The continuous law turns out to be (affine) linear with user speed. The disparities in the powers allocated to various speeds, increases with the increase in the path-loss factor and (or) the cell size. Further the discrete power law converges to continuous one, as number of speed classes increases to infinity. The numerical experiments confirm a remarkable improvement in drop probability with
optimal power law, in comparison with the case where equal power is allocated to all the users.
The
 performance improves as the number of speed classes increases and best performance is obtained with continuous law.
However, a significant improvement is achieved with just \textit{two} user-speed classes.

The optimal power law is derived under the assumption of \textit{zero} \textit{interference} from other cells due to theoretical tractability. However, we built a system level small cell network simulator (SCNS) to test the performance of optimal power law and we included some test cases with interference.  The SCNS is made by emulating cars moving along a straight
road (with random speeds), and receiving (random amount of) service from a series of pico towers adjacent to the road.
This simulator assumes perfect transmission when one operates below the capacity (at that time and at that position). First, we validate SCNS in order to test the optimal power law. We notice that power law \textit{fails} only when the block/drop probabilities are large (of order $10^{-1}$). However, wireless networks seldom operate in such scenarios due to low efficiency.

SCNS is used extensively to test the applicability of the proposed power law. We consider different test cases spanning over various configurations, for example, with a common set of transmission rates, and or with different levels of interference and or with various user speed profiles etc.  We notice a significant improvement (most of the cases above 30 \% and upto 89\% in some cases) in performance with linear power law in comparison with the systems that allocate equal power to all users.
The improvement is significant even for the test cases that do not satisfy the assumptions required by theory, e.g., test cases considering the influence
of interference or non-uniform speeds etc.
Even in the presence of interference we noticed good improvement (some cases upto 70 \%).
Further, we briefly  discuss the applicability of SCNS to chose the optimal system parameters.


\textbf{Related work:} For an excellent survey on power control in wireless networks, readers are referred to \cite{Chiang_now} and references therein (e.g., \cite{Foschini_ITVT, Bertsekas, Goldsmith, Ulukus, Yates, Ephremides}).
Most of the existing algorithms focus on either optimizing the total power spent while maintaining QoS or optimizing the QoS under a power budget constraint. 
Further refinement on these algorithms are based on speed of the user (e.g., \cite{MAZ,RPK,HLee} are few of them).
These algorithms either vary a control parameter or a step size of the adaptive algorithm so that the algorithm is better suited to the user mobility. They explore the convergence issues associated with mobility patterns. While we consider a small cell network with one dimensional users where: a) the service rate is adapted based on the received SNR; and hence b) the rate varies in a periodic fashion as the distance between unidirectional user and the serving BSs changes periodically. Given such a system, we obtain the performance which is sensitive to the speed of the user due to frequent HOs. We come up with the classification of users based on their speed and power control, so as to improve the performance of the overall system which shares resources among all the users. Most of the existing algorithms attempt to overcome the hurdles created by difficult phenomenon such as high speed variations, whereas our fundamental goal is to design optimal systems based on these variations. We achieve this goal by deriving an optimal policy which can either be stored as a lookup table (in case of finite classes) or the formula can be remembered and system can allocate power using this stored information and the estimate of the user speed {(in case of continuous power control)}. One can use this in self organizing networks.

Authors in \cite{Perf} considered
uni-directional users travelling on a straight road, while deriving service from series of BSs.
In \cite{Perf}, it is assumed that the rate of communication can be changed continually that too with ``maximal" transmission rates (i.e., capacity). This
gives ``maximal" performance, which is not realistic. We also consider a similar uni-directional user scenario, but for a radically different
change in the selection of the transmission rates: communication can happen at one of the $N$ given choices of the transmission rates. At any given time, the rates chosen are less than the ``maximal" transmission rate (capacity) at that time.

Parts of this work were presented in \cite{Allerton}. In this paper, we generalized the set of rates and more importantly included the system level simulator.
The system should provide sufficiently large choices of transmission rates to obtain significant advantage of power law. The more the choices, the finer is the control and better would be the improvement in performance. In \cite{Allerton}, we assumed that system supports $N$ choices and the set of these choices is different for different speeds. In this paper, using the system level simulator, we also include a study with a common (possibly bigger) set of rate choices.
We then consider the influence of interference on the performance of the proposed linear law, by comparing it with the  system that  allocates equal power to all users.  We study this comparison using the system level simulator and perform detailed numerical analysis, which includes  random and time varying interference patterns resulting from randomly originated and moving users deriving service from the encountered  small cell base stations.    We  observe that the improvement in performance (though reduced once interference is considered),  is   significant even in the presence of  {\it{interference.}}

This paper is organized as follows. The system model is described in Section \ref{sec_model}, performance measures are derived in Section \ref{sec_userspeeddepanal}
and the optimal power law is obtained in Section \ref{sec_opt_power_law}. The simulator and the extensive numerical results are described in
Section \ref{sec_numericals}-\ref{sec_onerateset}. Section \ref{sec_conc} concludes the paper.

\section{System Model}
\label{sec_model}

\begin{figure}
\begin{minipage}{7cm}
\begin{center}
\includegraphics[width=9cm,height=12cm]{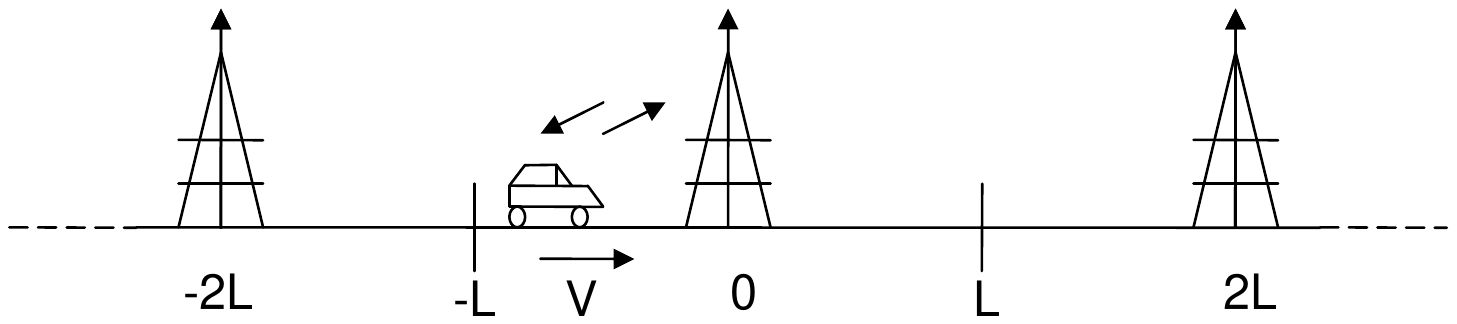}
\vspace{-98mm}
\caption{User moving in a car while deriving service. \label{Fig_OneDimCell}}
\end{center}
\end{minipage}
\hspace{2cm}
\begin{minipage}{7cm}
\begin{center}
\includegraphics[width=7.5cm,height=12cm]{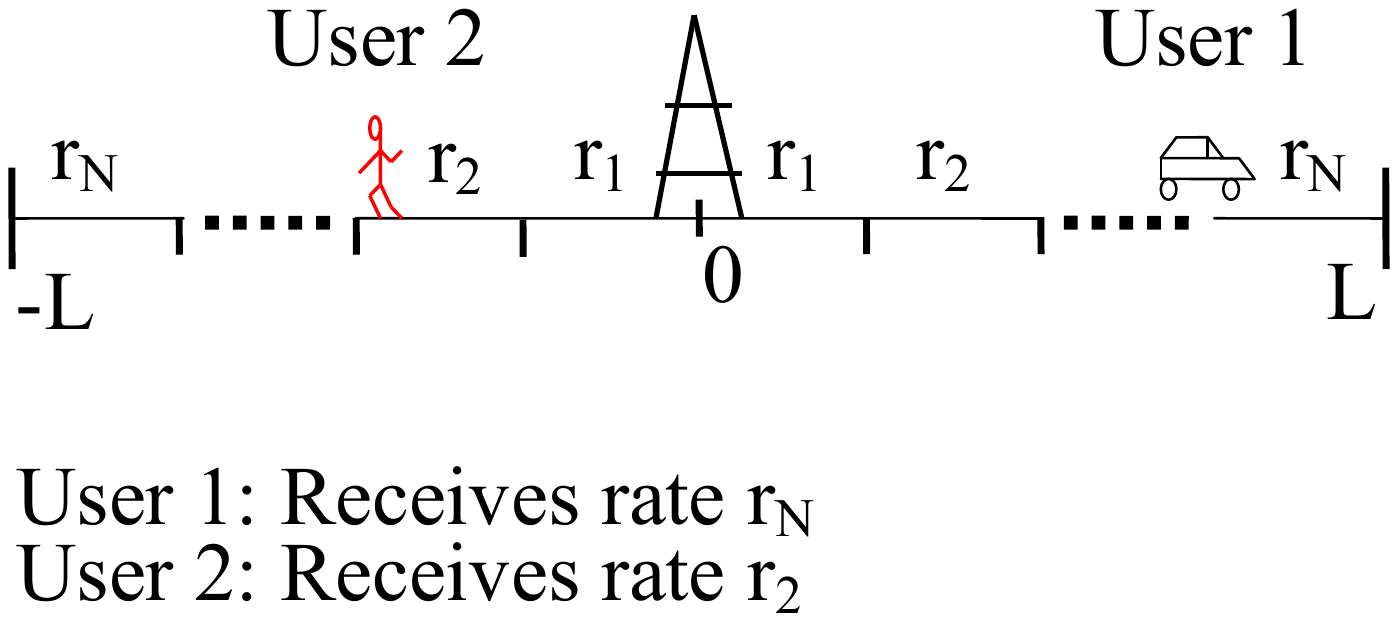}
\vspace{-95mm}
\caption{Rate partitioning and user's movement.\label{Fig_OneDimCellr}}
\end{center}\end{minipage}
\end{figure}

We consider users moving in a fixed direction.
The users are moving in one direction (in a one dimensional line $[-D,D]$) and at high speeds, which vary negligibly during the call. One such example (see Figure \ref{Fig_OneDimCell}) is when user, driving in a car, derives its service from portable BSs which are installed on street infrastructure (like lamp posts). The distance between successive BSs is $2L.$
 The users can move in one of the two directions with equal probability{, i.e., with half probability.
We assume symmetry in both the directions and hence any performance (e.g., busy probability, drop probability etc.), conditioned on the direction of the user, will be same
for both the directions. Thus, } without loss of generality we only consider users moving from {\it left to right.}
We assume the channel to be weakly time-varying. In particular, we assume that the optimal power allocation is achieved before significant channel variations, as is customarily assumed in all power control schemes. 
The user moves with speed $V$, which is independent and identically distributed (IID) across various users.
We assume uniform arrivals in the system. In small cell networks, the (adjacent) BSs are reasonably close and hence one has to design carefully with sufficient reuse factor to avail the advantages of small BS separations. In this work, we indeed assume this is the case for theoretical results,  i.e., we assume no interference from the other cells. Many other techniques like interference cancellation, beam forming etc are used to effectively mitigate interference. Alternatively one can assume interference to be constant and our results follow under such assumption on interference. This assumption is valid when one gets interference from large number of sources. 
In the last section of the paper, we present  numerical results considering {\it{interference}}. We  observe  that the performance of the optimal policies,   derived  in this paper, is remarkable even in the presence of interference.



\noi{\bf Rate Regions:}
The cell is divided into
$2N$ disjoint segments (depending upon distance from BS\footnote{In small cell networks (transmission at small distances), distance based propagation losses would be sufficient{ for deciding the theoretical rate limits as well as the practical transmission rates.}}) and users within a segment are served with the same transmission rate. Let $\{{\mathcal A}_n\}_{n \in {\mathcal N}}$ represent these {segments} (see Figure \ref{Fig_OneDimCellr}):
{{\small
\begin{eqnarray}
\vspace{-13mm}
\label{Eqn_CalA}
{\mathcal A}_{n} & := &
\left [ \phi_{n-1}L, \phi_{n}L\right ] 1_{\{n > 0\}} + \left [\phi_{n}L , \phi_{(n+1)}L\right ] 1_{\{n < 0\}},
 \label{Eqn_CalN}
 \nonumber \\ {\mathcal{N}} &:= &
\{-N, \cdots, -1, 1, \cdots, N \} ,
\end{eqnarray}}}where the coefficients $\{\phi_n\}$ determine the partition of the cell, {with $\phi_n = -\phi_{-n}$ for all $n$}.
Segments ${\mathcal A}_{n}, {\mathcal A}_{-n}$ placed {at the same distance}{} on either side of BS (Figure \ref{Fig_OneDimCellr}) are served with common rate $r_{n}$ and these common rates decrease as the distance from BS increases.
 Let ${\mathcal R} := \{ r_1, \dots, r_N\}$ (decreasing set) represent the ensemble of all possible transmission rates. The service rate changes once the user moves from one region to another.
{We consider $\phi_0 :=0$ for notational simplicity, and there is actually no change in rate when users move from ${\mathcal A}_{-1}$ to ${\mathcal A}_{1}$.}

\noi{\bf Arrivals:}
There are two types of arrivals: 1) arrivals from external world (represented by subscript $e$ and this subscript is used only when there is ambiguity) modelled as Poisson arrivals with parameter $\lambda$;
2) HO arrivals (subscript $h$) modeled again as Poisson\footnote{This is a commonly made assumption, see \cite{Rappa,Dharma}.} arrivals is the sub-stream of external arrivals whose service is not completed in the previous cell. Rate of arrivals into the cell of interest $[-L,L]$ (referred as cell 0){ depends upon the cell dimension $L$ and this is shown by either $\lambda_L$ (for external arrivals) or $\lambda_{h;L}$ (for HO arrivals). For
external arrivals, we assume}\footnote{If the arrivals in the entire line segment $[-D, D]$ occur at rate $\lambda'$ those
in segment $[-L, L]$ occur at a smaller rate $\lambda_L = \lambda' Prob (\mbox{arrival in } [-L, L]) $. For the special case of uniform arrivals (i.e., arrivals landing uniformly in $[-D,D]$)
 $\lambda_L = \lambda L$.} $\lambda_L = \lambda L$ {
 while $\lambda_{h;L}$ will be derived in later sections. }
Every arrival brings along with it the marks $(\Phi, S)$, where $\Phi \in \StatSpc$ is the position of arrival with distribution
$\Pi := \{ \pi_n \}_n$ and $S$, the number of bytes to be transmitted, is exponentially distributed, i.e., $S \sim \mu exp^{-\mu t} dt$ for some $\mu > 0$.

\noi{\bf Resources:} A cell can attend $K$ parallel calls.
{The power per transmission $P$ can depend upon the speed of the user and we obtain the optimal power function given an average power constraint.}

\noi{\bf Set of transmission rates ${\mathcal R}$:} With speed based power control,
there would be significant variations in the received power.
One can consider different set of rates for different speeds to design an efficient system. Alternatively, one can consider a common set of supported rates which is sufficiently big and choose one among them based on the received signal.
 First alternative is discussed below, while the other scenario is explored in section \ref{sec_onerateset}.
As of now, we assume that the system has predefined rate regions which are common for all the speeds/transmit powers.
The (common) rate supported in any given rate region depends upon the transmit power and the farthest point of the region.
The capacity of the farthest point provides the best possible rate.

We use the following low SNR approximation of the theoretical (capacity) rate\footnote{
For unit noise variance, capacity equals
$\log (1 + SNR)$, where signal to noise ratio $SNR = P A$ and attenuation $A = 1_{\{ d \le d_0\}} + (d/d_0)^{-\beta} 1_{\{ d > d_0\}}.$ For low SNRs,
 $\log (1+ SNR) \approx SNR $ and hence capacity equals $P A$.}:

{\small \vspace{-4mm} \begin{eqnarray}
 r(d) := P \left ( 1_{\{d \le d_0\}} + r_0 \left |d\right |^{-\beta} 1_{\{d > d_0\}} \right ) \mbox{ with } r_0 = d_0^\beta. \label{Eqn_cap}
\end{eqnarray}}
Here, $r(d)$ is the rate at distance $d$, $d_0$ is a small lossless distance\footnote{{ Typically
$d_0$ is small and in this paper {\it we consider optimizing over cell sizes $L > d_0 N$}}. That is, we always consider distances $d > d_0$ and hence use
the simpler formula
$r(d) = r_0 P d^{-\beta} $. } while $\beta$ is the propagation co-efficient. The farthest user of ${\mathcal A}_n$, will be at distance $\phi_nL$. Hence the maximal transmission rate,
 that can be allocated, equals
 \begin{eqnarray}
\label{Eqn_rates}
 r_n = r \left (\phi_n L \right ) = {r_0 P}{} L^{-\beta} \phi_n^{-\beta}.
 \end{eqnarray}
 {\it Alternatively, if
the system under consideration can design modulation and or channel coding schemes so as to achieve (almost) $\nu$ percent of the theoretical rates where $\nu < 1$ is a fixed coefficient, then again the above rate structure is applicable (after absorbing $\nu$ into $r_0$ of (\ref{Eqn_rates})).
 } {Practically, one probably can support only $\nu$ percent of true rates $r_n$, but $\nu$ can be absorbed into $r_0$.}

{\noi}{\bf Handovers:}
When the user reaches the boundary $\{ |{\bf x} | = L\}$ the call is handed over to the neighbouring cell (if the call is not completed and free servers are available in the new cell).

\noi{\bf Information to initiate HO:}
 Every new connection requires $s_h$ extra bytes to be exchanged to initiate it. The effect of these bytes (on the system performance) for a new call will be negligible (as it would be once),
 however one needs to consider their effect on HO calls.
These bytes are usually very small in proportion to the actual information to be transmitted, i.e., $s_h \ll S$ with probability close to one.
In particular we assume that $\mu s_h \ll 1$.
 We also assume that {\it $s_h$ bytes are exchanged with probability very close to one (actually w.p.1), while user is in {the last region }{}$r_{-N}$. }

\section{User speed dependent Performance measures}
\label{sec_userspeeddepanal}

In this section we derive the performance metrics (e.g., load factor, busy probability, drop probability etc), which capture the trade off between HO losses and the service quality, for a given user speed profile. We eventually conclude this section by showing that the optimizers of load factor and busy probability are the same.

We consider non elastic traffic which comprises of users demanding immediate service. These users (e.g., voice calls) drop the call if it is not picked up immediately, i.e., if all the $K$ servers are busy (see \cite{Perf,OWiOpt}).
{ The probability that a call is not picked up immediately is called the {{\textit{Busy probability}}} $P_{Busy}$ and
the probability that a call that was picked up is ever dropped before completing its service is called the {{\textit{Drop probability}}} $P_{Drop}$.
The former is the probability that an incoming call is not picked-up, while the later is the probability that, a call that was picked-up, is dropped before service completion.}
 Both these performance metrics
depend upon HOs as well as the average service time and hence capture the required trade-off.

In this paper we work with metrics proportional to $P_{Busy}$, to be specific with load factor $\rho$ (equation (\ref{Eqn_rhosing}) derived in later sections).
{\it One can obtain $P_{Drop}$ as in \cite{Perf} and show that it can be approximately optimized by minimizing the load factor, but we omit this discussion here. Similarly in \cite{Perf}, it is shown (under certain conditions) that the expected waiting time of an elastic user (a user who can wait for its turn but demands high quality of transmission)
can be optimized by minimizing the load factor. We expect similar results to hold even in this case.}
{\it In all, we optimize the load factor $\rho$. }

Without loss of generality we consider cell 0, $[-L,L]$.
Let $\psi_n$ represent the probability that a call originated in rate segment $n$, ${\mathcal A}_n$, is completed before reaching boundary $\{L\}$ (one element set). The user is served at rate $r_{n}$ and hence a total of $(L/N)(1/V) r_{n}$ bytes are transferred before it crosses ${\mathcal A}_n$ and then served at rate
$r_{n+1}$ and so on till it reaches the boundary point $L$. Total number of bytes transferred during this transit equal
$\sum_{m = n}^N r_{m} L /(NV) $ and so the user will complete its call (i.e., transfer of $S$ (exponential) bytes of information) before leaving the cell with probability (conditioning on $V$):

{{\small \vspace{-6mm}
\begin{eqnarray}
\label{Eqn_psin}
\psi_n =Prob \left (S < \frac{ L}{N V} \sum_{m = n}^{N} r_{m} \right ) = 1 - E\left [ e^{ - \frac{\mu L \sum_{m=n}^N r_{m} }{N V}}\right ].
\hspace{-5mm}
\end{eqnarray}}
In general it is difficult to obtain the above Laplace transform and hence $\psi_n$. For example, when $V$ is uniformly distributed.
However, one can obtain a good approximation if we assume, $ V_{min} \le V \le V_{max}$, with
 $V_{max}$ close to $V_{min}$ and both away from 0.
When the high speed users are
partitioned to a large number of classes, each class will satisfy the above assumption.
Further when $\mu$ is small (large jobs) and/or $V_{min}$ is large (as with high speed users of this paper) one can use
the approximation,
 $e^{-x} \approx 1 - x$, and then

{\small \vspace{-4mm}\begin{eqnarray}
\label{Eqn_psin1}
\psi_n \approx \frac{\mu L \sum_{m=n}^N r_m }{N} E[1/V].
\end{eqnarray}}
On the other hand, one can approximate $1-\psi_n$ directly with $1$ as $\psi_n$ in (\ref{Eqn_psin1}) is very small in comparison with 1.
{Let $P_{e,ho}$ ($P_{h, ho}$) represent the probability that a new or external (handover) call is handed over (again) to the neighbouring cell.} Note that we are modelling the HOs also as Poisson arrivals. A new call can arrive in any segment ${\mathcal A}_n$ with probability $\pi_n$ while a HO call always occur at $-L$, i.e., in rate region ${\mathcal A}_{-N}$ (calls are from left to right).
Then by unconditioning (w.r.t. the event of arrival being in ${\mathcal A}_n$),
{\vspace{-5mm}

\small \begin{eqnarray}
\label{Eqn_Peho}
P_{e,ho} &=& 1- \sum_{n=-N}^N \pi_n \psi_n \approx 1 - P L^{1-\beta} E[1/V] C_{e,ho} \nonumber \\
C_{e,ho} &:=& \frac{\mu}{N} r_0 \sum_{n = -N}^N \pi_n \sum_{m=n}^{N} \phi_m^{-\beta}.
 \end{eqnarray}}
In the above $C_{e,ho}$ is an appropriate constant and is obtained by substituting $r_n$ from (\ref{Eqn_rates}).
{\it Every HO call needs exchange of extra $s_h$ control bytes } and an HO call arrives only in ${\mathcal A}_{-N}$.
 Thus using similar logic ($C_{h,ho}$ is another constant like $C_{e,ho}$), $P_{h,ho} = 1- \psi_{-N,h}$ and so:

{\vspace{-5mm} \small \begin{eqnarray}
\label{Eqn_Phho}
P_{h,ho}& = & 1- Prob \left (S+s_h < \frac{ L}{N V} \sum_{m = -N}^{N} r_m \right ) \nonumber \\
&& \hspace{-10mm} \ \ \approx \ \ 1- \left (C_{h,ho} P L^{1-\beta} E[1/V] - \mu s_h \right ) \mbox{ with } \nonumber \\
C_{h,ho} &:= & \frac{\mu}{N} r_0 \sum_{n = -N}^N \phi_n^{-\beta}.
\end{eqnarray}}

 \subsection{Expected Service time}
The expected amount of time for which a user is served in a cell is the expected service time. 
{
Let $b_n$ represent the average time for which the service is received in cell 0, given call originated in region ${\mathcal A}_n$. When {\it high speed users travel in small cells}, with high probability, a call is not completed in one cell and hence one can approximate $b_n$
 with the time taken to reach the boundary:
$${
b_n = \sum_{k \ge n} (\phi_{k+1}- \phi_k) L E\left [ \frac{1}{V} \right ]} \mbox{ = } L \left ( \phi_N-\phi_n \right ) E\left [ \frac{1}{V} \right ].
$$}
{
Recall that $\phi_n = -\phi_{-n}$.} The service time of a new call $(b_{e})$ and that of a handed over call ($b_{h}$) on average equals (by unconditioning)\footnote{The time taken for exchange of $s_h$ HO bytes has to be included in the time of the cell utilized by the user and hence
the (HO) service time $b_h$. However, the rate of HOs is high and we are approximating $b_h$ by the average time taken to traverse the entire cell, $b_h$ does not change with $s_h$.
}:
{\small
\begin{eqnarray}
\label{Eqn_be}
b_e &=& \sum_{i = -N}^N b_n \pi_n = C_{b,e} L E[1/V] \mbox{, } { C_{b,e} := \sum_{n=-N}^N \pi_n (\phi_{N}- \phi_n)} \nonumber \\ b_{h} &=& b_{-N} = C_{b,h} L E[1/V]
 \mbox{, } \hspace{10mm} C_{b,h} := 2. \hspace{-5mm}
 \end{eqnarray}}
 The expectation in $E[1/V]$ is with respect to the distribution of the user speeds. However, the corresponding distribution in $b_h$ (the average service time for HO users)
 is the speed distribution of the HO users, which may not be the same as that of the new users. The higher speed users are handed over more frequently than the lower speed ones. So, the distribution of the HO user velocities can be different. In \cite[Appendix B]{Perf}, it is shown that the HO user speed distribution converges to new-user speed distribution as the cell size reduces to zero, when the new-user speed distribution is uniform. {\it In exactly similar lines one can show the same result for any arbitrary new-speed distribution, when it has a density.
 Thus we approximate the HO user speed distribution with the new-user speed distribution.}
Further and more importantly, this approximation is more accurate as the support ($V_{max}-V_{min}$) decreases.
Or equivalently the approximation is more accurate,  when
 the number of velocity based user classes, considered in section \ref{sec_opt_power_law}, increases.

Performance metrics obtained in the remaining part of this section, section \ref{sec_userspeeddepanal}, are derived in similar way as in \cite{Perf} and \cite{OWiOpt}.
These derivations are obtained
using the stochastic equivalence of calls going out of and coming into cell 0 (details in \cite{Perf,OWiOpt}).
We present the derivations briefly and more details can be found in the two papers. These results are useful in deriving optimal power law in {section \ref{sec_opt_power_law}}.

\subsection{HO Arrival rate}{
 We obtain the rate, $\lambda_{h;L}$, at which HOs occur. Let $\lambda_L $ represent the fraction of the new calls that arrive in the cell of interest $[-L, L]$ which equals $\lambda L$ for uniform arrivals (for appropriate $\lambda > 0$). A fraction of the new arrivals as well as HO calls get converted (again) to HO calls. The calls that have not finished their service before reaching the boundary are exactly this fraction, whose value is given by $P_{e,ho}$ and $P_{h,ho}$ respectively for new arrivals and the HO calls. By memory less nature of $S$ (the bytes to be transferred), we have a stochastic equivalence between the calls entering and leaving the cell (see \cite{Perf}, \cite{OWiOpt}). Using this, $\lambda_{h;L}$ satisfies the fixed point equation,
$
\lambda_{h;L} = \lambda_L P_{e,ho} + \lambda_{h;L} P_{h,ho}.
$ Hence, from equations (\ref{Eqn_Peho}), (\ref{Eqn_Phho})}
{

\vspace{-4mm}
\small
$$
{}
\lambda_{h; L} = \frac{\lambda_L P_{e,ho}}{1 - P_{h,ho}} = \lambda L \frac{ 1 - P L^{1-\beta} E[1/V] C_{e,ho}}{P L^{1-\beta} E[1/V] C_{h,ho} - \mu s_h}.
$$\hspace{-1mm}}
\subsection{Overall expected service time} ({On} considering {both} HO and new arrivals) {
}
{\small
$$
{\bar b} = \left ( \frac{\lambda_L}{\lambda_L + \lambda_{h; L}}b_{e} + \frac{\lambda_{h;L}}{\lambda_L + \lambda_{h; L}} b_{h} \right ).
 $$}
 By ergodicity,
 the probability that a given call is new, equals the ratio of the arrival rate of new or external calls and the arrival rate of all the calls and hence the above result.
\subsection{Load factor} The product of the average service time and the arrival rate gives the rate at which the overall load is arriving into the system and this product divided by the number of servers is called \underline{load factor}, which is given by:
\vspace{-1mm}
$$
\hspace{15mm}
\rho = (\lambda_L + \lambda_{h;L}) {\bar b} / K.
\vspace{-2mm}
$$
On simplification,

\vspace{-4mm}
{\small \begin{eqnarray}
\label{Eqn_rhosing}
\rho =
\frac{\lambda L^2 E[1/V]}{K} \left (C_{b,e} + C_{b,h} \frac{1 - P L^{1-\beta} E[1/V] C_{e,ho}}{ P L^{1-\beta} E[1/V] C_{h,ho} - \mu s_h}\right ).
\end{eqnarray}
}
\vspace{-2mm}
\subsection{Busy Probability}
{
A small cell catering to
non elastic traffic can be modeled by an M/G/K/K queue (as in \cite{Perf}). By Erlang's loss formula, busy probability}
{\small
 \begin{eqnarray}
 \label{Eqn_Pbusy}
 P_{Busy} = \frac{\rho^K / K! }{ \sum_{k=0}^K \rho^k / k!}. { \mbox{ It depends on $L$ only via $\rho$ (see \cite{LongTR})} \mbox{ and so}}
 \end{eqnarray} }
{
Thus from above, the busy probability depends upon $L$ only via $\rho$ and both are differentiable in $L$. By differentiating twice one can
 immediately obtain the following result (see \cite[Theorem 5]{Perf} for similar details). }{}
 \begin{lem}\label {Lemma_BusyRho}
 Optimizers of $\rho$ and $P_{Busy}$ are same, i.e.,
$$L^*_\rho := \arg \inf_{L} \rho = \arg \inf_{L} P_{Busy} (L) =: L^*_{P_{Busy}} .$$
 \end{lem}
By Lemma \ref{Lemma_BusyRho}, the cell size optimizing $P_{Busy}$ is the same as that optimizing the load factor.
In a similar way the power control affects busy probability only via the load factor $\rho$.
Hence we optimize load factor ($\rho$) in subsequent sections.

\section{Optimal power law}
\label{sec_opt_power_law}

In this section, we derive optimal power law for the given cell size $L$ and power constraint ${\bar P}$. Subsequent sections discuss the optimal power law for finite and infinite number of user classes.

\subsection{Finite number of User classes}

The users are divided into different classes based on their speeds. 
 Divide the (speed) interval $[V_{min}, V_{max}]$ into $I$ disjoint intervals $\{{\cal I}_i \}_{i \le I}$ and
 classify users into one of the $I$ classes based on their speed. 
 From (\ref{Eqn_rhosing}), load factor $\rho$ depends upon the user speed profile only via $E[1/V]$ and so we are interested
in the conditional expectation
\vspace{-1mm}
$$
\hspace{5mm}
\Upsilon_i := E\left [ \left . \frac{1}{V} \right | V \in {\cal I}_i \right ].$$
Let $p_i := Prob (V \in {\cal I}_i)$ represent the probability of class $i$ and $P_i$ is the transmit power allocated to class $i$. {

HO rates of the different user classes can be calculated using earlier logic: 

{\small \[
\lambda_{h;L,i} = \lambda_{L,i} \frac{P_{e,ho,i}}{1 - P_{h,ho,i}} = \lambda L p_i \frac{ 1 - P_i L^{1-\beta} \Upsilon_i C_{e,ho} }{ P_i L^{1-\beta} \Upsilon_i C_{h,ho} - \mu s_h}. \]}
 Similarly the expected service time for different user classes can be obtained and then the} overall load factor
 $\rho$ simplifies\footnote{Load factor $\rho$ equals the product of total average arrival rate (${\bar \lambda} := \lambda_{L} + \sum_i \lambda_{h;L,i}$) and average service time divided by $K$.
An arrival turns out to be a new arrival with probability $ \lambda_L/{\bar \lambda}$, it turns out to be an HO arrival of class $i$ with probability
$\lambda_{h;L,i}/{\bar \lambda}$ and so average service time equals $\left ( \lambda_L b_e + \sum_i \lambda_{h;L,i} b_{h,i} \right )/{\bar \lambda} .$}:
\begin{eqnarray}
 \rho (L, {\bf P}) &=& \frac{1}{K} \left (\lambda_L b_e + \sum_{i} \lambda_{h;L, i} b_{h,i} \right ) =
 \frac{\lambda L^2}{K} \sum_i p_i \Upsilon_i \left ( C_{b,e} + C_{b,h} \frac{1 - \delta_i C_{e,ho}}
{ \delta_i C_{h,ho} - \mu s_h} \right ) \hspace{-1mm}\label{Eqn_rho}
\end{eqnarray}
where $\delta_{i}:= P_i\Upsilon_iL^{1-\beta}$.
We are interested in power control ${\bf P} := \{P_1, \cdots, P_I\}$ which minimizes $\rho$ while the average power satisfies following constraint:
$$\sum_i p_i P_i \le {\bar P}.
$$

Assuming that $\bar{P}$ is sufficiently large to ensure useful communication (see \ref{powerassumption}), one can solve above optimization problem and we obtain:

\begin{thm}
\label{Thrm_DiscretePowerLaw}
Assume $C_{h,ho} - \mu s_h C_{e,ho} > 0$ and assume that the following matrix is positive definite,
$$
 {\cal P}_V := \left [
\begin{array}{lllll}
p_1 + \frac{p_1^2}{p_I} & \frac{p_1 p_2 }{p_I} & \cdots & \frac{p_1 p_{I-1}}{p_I} \\
 \frac{p_1p_2}{p_I} & p_2 + \frac{ p_2^2 }{p_I} & \cdots & \frac{p_2 p_{I-1}}{p_I} \\
 & & \vdots \\
 \frac{p_1p_{I-1}}{p_I} & \frac{p_{I-1} p_2 }{p_I} & \cdots & p_{I-1}+\frac{ p_{I-1}^2}{p_I} \\
\end{array}
\right ] > 0.
$$
Then, the power control that minimizes $\rho$, given by (\ref{Eqn_rho}), while satisfying the average power constraint equals:

{\small \begin{eqnarray}\boxed{
\label{Eqn_OptPower}
P_i^* (L; {\bar P})
 = {\bar P} + \frac{\mu s_hL^{\beta-1}}{ C_{h,ho}} \left (\frac{1}{\Upsilon_i} - \sum_{j} p_j \frac{1}{\Upsilon_j} \right ), \ \Upsilon_i := E\left [ \left . \frac{1}{V} \right | V \in {\cal I}_i \right ]. \hspace{2mm}}
\end{eqnarray}}
\end{thm}
 {\bf Proof:} is provided in Appendix B. \hfill{$\Box$}

From (\ref{Eqn_Peho})-(\ref{Eqn_Phho}) $C_{h,ho} \ge C_{e,ho}$ and from our earlier assumptions, $\mu s_h \ll 1$. So, the first assumption would be trivially satisfied. The second assumption is often satisfied, for example when users are classified into $I$ uniform classes, i.e., when
$$
 {\cal I}_i = V_{min} + \left [ \frac{i-1}{I}, \frac{i}{I} \right ] (V_{max}-V_{min}) \hspace{1mm} \mbox{ and so } p_i = \frac{1}{I} ~\forall~ i.
$$


We obtain the following interesting properties of the optimal power division (from
(\ref{Eqn_OptPower})):
1) allocated power increases with the speed range of users in a class (users with lower $\Upsilon_i$ get higher power), which in turn implies an increased virtual cell;
2) the disparity in the allocated powers between different classes increases with cell size, the disparity in speeds and the path-loss factor. {\it Thus
from
(\ref{Eqn_OptPower}) in a medium with large (path) losses, one needs to allocate larger power to high speed users in order to improve the overall system performance. Intuitively otherwise, large speed users hold on to the system resources for longer time periods which can deteriorate the performance of the low speed users. And the same applies to the case when the system has to support wide range of user speeds ($V_{max}- V_{min}$).}

\subsection{Continuous optimal power law}
When one has an accurate estimate of the velocity, it might be optimal to do finer power control, i.e., transmit power is varied continually based on the precise value of velocity. Further, as mentioned in the previous sections, the modelling in-accuracies (while obtaining HO rates, HO velocity distribution etc.) mitigate as the number of classes increase. In this continuous case, we expect to model more accurately.

We assume that user velocity is a continuous random variable with density $p_V (v)$ and with support in the range $[V_{min}, V_{max}]$ for some
$0 < V_{min} < V_{max} < \infty$. For example for uniform case, $p_V(v) \equiv 1/(V_{max}-V_{min})$.

We obtain an optimal power law, which is a function of velocity $v$ and optimizes the load factor $\rho$. By power law we meant a function $P(.)$ which maps velocity $v$ to a positive real number, that represents the power allocated to user travelling with speed $v$.
 We are only interested in those functions,
whose average is constrained by ${\bar P}$, i.e., we are interested in policies, $\{P(.)\}$, that satisfy:
$$
\hspace{10mm}
\int_{V_{min}}^{V_{max}} P(v) p_V (v) dv \le {\bar P}.
$$

Let $B$ represent the random service time (the time for which the cell resources are utilized by the user).
The conditional average service times $b_e (v)$, $b_h (v)$, the probabilities of service not getting completed $P_{e,ho} (v)$, $P_{h,ho}(v)$ and the HO rates $\lambda_{h;L} (v)$ can be computed by conditioning on user velocity $V$ as in the previous section. These expressions depend upon the power policy $P(.)$. It is easy to see that these conditional expressions will be same as in the previous section except that $\Upsilon_v = E[1/V | V \in (v-dv, v+dv) ]$ (conditioned that $V$ is in an infinitesimal interval around $v$) has to be replaced by $1/v$, i.e., for example:
 \begin{eqnarray*}
 b_e (v) &=& \frac{C_{b,e} L}{v} \mbox{, } \ \ P_{e,ho} (v) = 1 - \frac{P(v) L^{1-\beta}C_{e,ho}}{ v } \\
 b_{h} (v) & =& \frac{C_{b,h} L }{v} \mbox{, } \ \ P_{h,ho}(v) = 1- \frac{P (v) L^{1-\beta} C_{h,ho} }{v} + \mu s_h.
 \end{eqnarray*}

 \subsubsection*{Overall arrival rate}
 By conditioning and un-conditioning on velocity $V,$ the overall arrival rate including the HOs, equals ($E$ is the expectation with respect to $V$):
\begin{eqnarray}
\label{Eqn_Ctsbarlambda}
{\bar \lambda} = E [ \lambda_L (V) + \lambda_{h;L} (V) ] = \lambda L E \left [ 1+ \frac{P_{e,ho}(V)}{1 - P_{h,ho} (V)} \right ].
\end{eqnarray}
\subsubsection{Average of the overall service time $B$}
The expected value of $B$ is obtained in \ref{sec_expectedB} and it equals:
{\small \begin{eqnarray}
\label{Eqn_Ctsbarb}
{\bar b} &=& \frac{ E \left [ b_e (V) + b_{h} (V) \frac{P_{e,ho}(V)}{1 - P_{h,ho}(V)} \right ] }
{ E \left [ 1+ \frac{P_{e,ho}(V)}{1 - P_{h,ho} (V)} \right ]}\\ && \hspace{-12mm}=
 \frac{L}{ E \left [ 1+ \frac{P_{e,ho}(V)}{1 - P_{h,ho} (V)} \right ]}
\int_{V_{min}}^{{V_{max}}} \left ( C_{b,e} + C_{b,h} \frac{1 - \frac{P(v) L^{1-\beta}C_{e,ho}}{ v }} {\frac{P (v) L^{1-\beta} C_{h,ho} }{v} - \mu s_h} \right ) \frac{p_V (v)}{v} dv. \nonumber
\end{eqnarray}}
 And so the load factor

 {\small
\begin{eqnarray}
\rho =
 \frac{\lambda L^2}{K} \int_{V_{min}}^{{V_{max}}} \left ( C_{b,e} + C_{b,h} \frac{1 - \frac{P(v)C_{e,ho}}{ L^{\beta-1} v }} {\frac{P (v) C_{h,ho} }{L^{\beta-1} v} - \mu s_h} \right ) \frac{p_V (v)}{v} dv. \label{Eqn_ctsrho}
\end{eqnarray}}
We are interested in power law $\{ P^*(.) \}$ which minimizes the load factor $\rho$ under the average power constraint.
One can rewrite:
$$
 \frac{1 - \frac{P(v)C_{e,ho}}{ L^{\beta-1} v }} {\frac{P (v) C_{h,ho} }{L^{\beta-1} v} - \mu s_h} =
 - \frac{C_{e,ho}}{ C_{h,ho}} + \frac{1 - \mu s_h \frac{C_{e,ho}}{ C_{h,ho}} } {\frac{P (v) C_{h,ho} }{L^{\beta-1} v} - \mu s_h}.
$$
We again assume, $C_{e,ho } \mu s_h < C_{h,ho}$, this makes the numerator of the second factor positive and after leaving out the constants the optimal power law is given by:
\begin{eqnarray}
P^*(.) = \arg \inf_{P(.) } \int_{V_{min}}^{{V_{max}}} \frac{1 } { P (v) C_{h,ho} - v L^{\beta-1} \mu s_h} p_V (v) dv \nonumber
\end{eqnarray}\begin{eqnarray}
 \mbox{subject to }
\int_{V_{min}}^{{V_{max}}} p_V (v) P(v) \le {\bar P}. \hspace{20mm}
\end{eqnarray}
With $\varrho$ representing the Lagrange multiplier we minimize:
\begin{eqnarray}
 \int_{V_{min}}^{{V_{max}}} \left ( \frac{1 } { P (v) C_{h,ho} - v L^{\beta-1} \mu s_h} - \varrho (P(v) - {\bar P}) \right ) p_V (v) dv. \nonumber
\end{eqnarray}
This is a state independent optimal control problem, for any given $\varrho$ and can be solved using Hamilton-Jacobi-Bellman (HJB) equation (see
for e.g., \cite{BasarOlsder99,Fleming}):
\begin{eqnarray}
\label{Eqn_HJB}
\frac{d U }{d v} = p_V (v) \inf_P \left \{ \frac{1 }{C_{h,ho}P - \mu s_h L^{\beta -1} v} - \varrho P
 \right \}
\end{eqnarray}
 where $U$, the value function, is given by:
 {\small
 $$
 U( v) := \inf_{ P (.)} \int_{v}^{{V_{max}}} \hspace{-2mm} \left [ \frac{1 } { P (v) C_{h,ho} - v L^{\beta-1} \mu s_h} - \varrho P(v) \right ] p_V (v) dv. \nonumber
 $$ }
 The value function $U$ satisfies the HJB equation (\ref{Eqn_HJB}) when it is continuously differentiable (e.g., \cite{BasarOlsder99,Fleming}) and we will see that this indeed is the case.
The optimal control $P^*()$ equals the minimizer of the optimization in the right hand side of the HJB equation (\ref{Eqn_HJB}) (see for example \cite{BasarOlsder99,Fleming}). This can be computed easily
for any $v$ and
$$
P^* (v) = \frac{1}{C_{h,ho}}\mu s_h L^{\beta-1} v + \sqrt{\frac{1}{C_{h,ho} \varrho}}.
$$ The average power equals:
$$
\int_{V_{min}}^{V_{max}} P^* (v) p_V (v) dv = \frac{1}{C_{h,ho}}\mu s_h L^{\beta-1} E[V] + \sqrt{\frac{1}{C_{h,ho} \varrho}}.
$$ Equating the above to average power constraint ${\bar P}$, we obtain the optimal Lagrange multiplier $\varrho$ and optimal power law:
\begin{thm}
\label{Thm_cts_law}
The power function $P^*(.)$ that optimizes $\rho$ given by (\ref{Eqn_ctsrho}), while satisfying the average power constraint equals:
\begin{eqnarray}
\label{Eqn_cts_opt_power}
P^*(v) = {\bar P} + \frac{1}{C_{h,ho}}\mu s_h L^{\beta-1} \left ( v - E[V] \right ). \hspace{5mm }\Box
\end{eqnarray}
\end{thm}
 {\bf Remarks:} From (\ref{Eqn_cts_opt_power}), we notice that the optimal power law is linear in the speed. In fact even from discrete case (\ref{Eqn_OptPower}),  it is proportional to $1/\Upsilon_i $ which approximately equals a speed $v \in {\cal I}_i $ if  interval ${\cal I}_i $  is sufficiently thin. This is surprising  for us, as we anticipated that the power law will be scaled by factors proportional to path-loss factor $\beta$.  We reconfirmed using the small cell network simulator built in section \ref{sec_SCNS} that, the linear power law indeed performs superior in comparison to many other functions. We compared linear power law (\ref{Eqn_cts_opt_power}) with  power functions that are square or square root of the velocity ($P(v) \propto v^2$ or $\sqrt{v}$) or  functions  such that  
($P(v) \propto v^\beta$ or  $1/v^{\beta}$) etc., and observed that   the former  provides the best performance. 
However the coefficients of the linear power law depend upon the cell size $L$, path-loss factor $\beta$, amount of extra information required at each hand over $s_h$ etc.,  also as in discrete case. For example, higher the path-loss factor and or larger the cell size is, larger is the disparity in the powers allocated to various speed users.

One can easily verify that the continuous power law (\ref{Eqn_cts_opt_power}) equals\footnote{As one increases the number of user classes $I$ to infinity,
the conditional expectation $\Upsilon$ converges to $1/V$ while
the sum for any $I$ equals, $\sum_i p_i \Upsilon_i = E[1/V]$, the unconditional expectation.} the limit of the discrete optimal power policy (\ref{Eqn_OptPower}).
Thus we obtained optimal power law via two different methods for continuous control and the two methods are resulting in the same solution.
Also, the optimal power increases linearly with the speed $v$.

The load factor at optimal power law parametrized by threshold ${\bar P}$ equals:

{\small
\begin{eqnarray} \rho^* (L; {\bar P}) &=& \frac{\lambda L^2}{K} \int_{V_{min}}^{{V_{max}}} \left ( \frac{C_{\rho,1}}{v} +
 \frac{C_{\rho,2} }{{\bar P} L^{1-\beta} C_{h,ho} - E[V] \mu s_h}
 \right ) p_V (v) dv \label{Eqn_Optrho}
\\
&=& \ \frac{\lambda L^2}{K} \left ( C_{\rho,1} E\left [\frac{1}{V} \right ] + \frac{C_{\rho,2} }{{\bar P} L^{1 -\beta} C_{h,ho} - E[V] \mu s_h} \right )
 \mbox{ with}
\nonumber \\
\nonumber
&& \hspace{-12mm}
 C_{\rho,1} := C_{b,e} - \frac{C_{b,h} C_{e,ho}}{C_{h,ho}}, \mbox{ and }
 C_{\rho,2} := C_{b,h} \left ( 1 - \mu s_h \frac{C_{e,ho}}{ C_{h,ho}} \right ). \nonumber \hspace{-5mm}
\end{eqnarray}}

\section{Beta+ scaling and Optimal cell size }
\label{sec_opt_cell_size}

In \cite{OWiOpt}, while studying the randomly wandering users, it is shown that the performance degrades monotonically with cell size unless the power per transmission is boosted via beta+ scaling
(i.e., $P \propto L^{\beta + \gamma}$ for some $\gamma > 0$). In Appendix B, we show that the uni-directional users of this paper also require beta+ scaling.

In this section we obtain optimal cell size one for each $\gamma$, when the power is boosted by beta+ scaling (we actually boost the average power constraint) and when the power is allocated to users according to the optimal law (\ref{Eqn_cts_opt_power}).
This means we have an $L$ dependent power constraint i.e., ${\bar P} = {\tilde P} L^{\beta+\gamma}$ for some
${\tilde P}, \gamma > 0$ when obtaining the optimal power control for each $L$ and we assume ${\tilde P}$ is sufficiently large to ensure useful communication:
$$
L^* = \arg \inf_{\left \{ L \right \}} \rho^* \left (L; {\tilde P}L^{\beta + \gamma} \right ),
$$ with $\rho^*$, the load factor at optimal power law is given by (\ref{Eqn_Optrho}). However beta+ scaling implies increasing total power used by the system (the number of cells in the system is proportional to $1/L$ and hence the total power is proportional to ${\tilde P}L^{\beta+\gamma}/L$) and hence we consider the following joint cost to obtain the optimal cell size ($\omega_P$ is the weight given to power):
\begin{eqnarray}
\label{Eqn_jointcost}
L^*_{\hbar} &=& \arg \inf_L \hbar(L) \ \ \mbox{ with } \nonumber \\
\hbar(L) &:=& \rho^* \left (L; {\tilde P}L^{\beta + \gamma} \right ) + \omega_P {\tilde P} L^{\beta + \gamma -1}
\nonumber \\ &=&
 L^2 C_{\hbar,1} + C_{\hbar,2} \frac{L^2}{L^{\gamma+1} - C_{\hbar,3}} + C_{\hbar,4} L^{\beta + \gamma -1},
\end{eqnarray} where the constants $C_{\hbar,1}$ to $C_{\hbar,4}$ can easily be defined using (\ref{Eqn_Optrho}). It is usually difficult to obtain the closed form expressions for the optimizer, alternatively we obtain it via numerical examples in the next section for some case studies. One can get closed form expression for some special cases, either by equating the derivative to zero or by inspection. One can easily see that $L^*$ equals trivial $Nd_0$ if $\gamma = -1$ (as then $\rho$ is monotonically
increasing in $L$ for any $\beta \ge 2$). When $\beta = 2$ and $\gamma = 1$ we have the following (obtained using the first and the second
derivatives):
\begin{lem}
\label{Lemma_OptCellSiz}
For $\beta = 2$, $\gamma = 1$ and when $C_{\rho, 1} > 0$ the cell size that optimizes the joint cost $\hbar$ is given by (\ref{Eqn_jointcost}) equals :
\begin{eqnarray*}
L^* = \sqrt{ \frac{\sqrt{\mu s_h E[V]}}{{\tilde P} C_{h,ho}} \left ( \sqrt{ \frac{C_{\rho,2}}{C_{\rho,1} E[1/V] + \frac{\omega_P {\tilde P} K }{\lambda} } } + \sqrt{E[V] \mu s_h } \right ) }. \hspace{2mm} \Box
\end{eqnarray*}
\end{lem}
From the above lemma, we obtain the following general characteristics of the optimal cell size. Some of these conclusions are similar to the ones made for randomly wandering users of (\cite{OWiOpt}).
\begin{enumerate}[a)]

\item Larger the range of velocities (larger $E[V]$ or smaller $E[1/V]$), larger is the optimal cell size.
Basically the fraction of time for useful communication reduces with large speeds.

\item The smaller is the constraint on the power (smaller $\omega_P$) larger is the optimal cell size.

\item Smaller is the information required for HO in comparison with the information to be communicated (smaller $s_h$ or smaller $\mu$)
the smaller is the optimal cell size. As then the trade-off (between HOs and better service times) mentioned in the introduction biases more towards better service times and hence it is advantageous
to use smaller cell sizes.

\end{enumerate}
Though the above discussion is specific to one particular example, one can obtain similar characterizations for all the general cases via numerical examples.

\section{Small cell network simulator}
\label{sec_SCNS}

We built a small cell network simulator (SCNS) for elaborate study of the system under consideration. We validate the SCNS before using it for case studies.
In all the test cases we compare the performance of the system using proposed linear power law with that of the systems which allocate equal power to all the users.
The simulator also enables us to study the test cases that do not satisfy the assumptions required by the theory.  In the next section, we consider two such important generalizations. The pathloss factor $\beta$ has been set from $2.1$ to $3.5$. We have also assumed that users are moving with a speed ranging from $20$ Kmph to $100$ Kmph. The range of the small cells ($2L$) has been set between $140$m to $160$m using a power between $0.7$ Watt to $1.6$ Watt which corresponds to a typical outdoor metro cell or pico cell.

The SCNS emulates cars moving along a straight road, and receiving services from a series of pico towers adjacent to the road. Without loss of generality, we consider cars moving only in one direction. The SCNS emulates the arrival of new users, service of the user by the nearest tower based on SNR, hand over of services between towers and departure of users from system once service is completed. The simulator also accounts for the data rates lost due to transfer of extra control information per HO. It 
updates the value of variables at regular time intervals of length $\delta$, where $\delta>0$ is sufficiently small, to achieve `almost' continuous time simulations. The cars leaving the last tower is tied to the first tower to ensure the stochastic equivalence of cars entering and leaving the system.

Before we proceed, we would like to clarify an important underlying assumption. In SCNS, we assume that there are no transmission errors when one transmits at rates below the SNR (the approximation for capacity). Here, we are estimating the SNR at any position and then choosing the best rate (among $\mathcal {R}$) below the SNR. In that sense, these are the results of partial simulations and complete simulations would incorporate channel coding, decoding transmission errors etc. Now, we discuss \textit{several stages} involved in building SCNS.
 \\
1) \textit{Deterministic variables:} A single tower with a single server and a single supported rate is built with all other variables (e.g., inter arrival time, service requests etc.) as deterministic. \\
2) \textit{Stochastic variables:} We continued with single tower and single server, but the variables are randomised one by one. We also consider multiple transmission rates. We run simulations for considerably long time to estimate the confidence interval for each case.
 \\
3) \textit{Multiple servers and small cell network:} Finally, we considered the case with multiple towers (each having multiple servers), with the road from final tower leading back to that of the first tower.
The performance measures are estimated, after running the simulations for even longer periods to ensure convergence.

\subsection{Validation of SCNS}
 In order to validate our results of the previous sections, we compare the theoretical analysis with simulation results obtained by SCNS. But prior to that, one needs to verify the SCNS itself. Towards this, we build the SCNS in stages, as mentioned above, and some easily computable performance metrics are computed at each stage and are verified against the estimates provided by SCNS.
We consider many intermediate measures for this validation purpose.
Then, the important performance measures ($P_{Busy}$ and $P_{Drop}$) are estimated using SCNS and are compared with the theoretical ones. We also discuss the optimality of power law.

The results of SCNS are verified using corresponding theoretical expressions first under deterministic setting (when job sizes, user speed etc are considered constants) and then for stochastic system setting.  Then  the experiments are performed for small cell network with multiple servers. Theoretical performance measures derived in Section \ref{sec_userspeeddepanal} are applicable for this system under following assumptions: a) the values of $P_{e;ho}$ and $P_{h;ho}$ are  close to 1;
b) The values of busy and drop probabilities are small; and 
c) The values of $V_{min}$ and $V_{max}$ are far away from zero.  These assumptions are easily satisfied by almost all practical scenarios (to provide reasonable QoS) and we also consider case studies satisfying these assumptions.

We compare the values of $P_{e,ho},~P_{h,ho},~b_e,~b_h$ in Table \ref{netwok_validation}.
First three sub-columns mention the configuration, next  4 sub-columns provide the intermediate performance measures estimated using SCNS and the last four display the corresponding theoretical performance. 
 It is clear that there is a good match (around 1 to $5\%$ normalized difference) between the theoretical and the values estimated using SCNS, under different configurations of power ($P$), transmission rates set (${\mathcal{R}}$) and number of BSs ($n$). This completes the validation of SCNS.

\begin{table}[h]
\begin{center}

\begin{tabular}{|l|l|l|l|l|l|l|l|l|l|l|}
\hline
\multicolumn{3}{|c|}{{ Configuration}} & \multicolumn{4}{c|}{{ SCNS Results}} & \multicolumn{4}{c|}{{ Theoretical Results}} \\ \hline
$P$ & ${\mathcal{R}}$ & $n$ & $P_{e,ho}$ & $P_{h,ho}$ & $b_e$ & $b_h$ & $P_{e,ho}$ & $P_{h,ho}$ & $b_e$ & $b_h$ \\ \hline
1.2&[0.6 0.001]&3&0.94&0.88&2.67&5.27&0.94&0.87&2.84&5.68\\ \hline
1.2&[0.6 0.001]&5&0.94&0.89&2.68&5.29&0.94&0.87&2.84&5.68\\ \hline
1&[0.6 0.001]&3&0.94&0.89&2.68&5.29&0.94&0.88&2.84&5.68\\ \hline
1&[0.6 0.001]&5&0.94&0.89&2.68&5.29&0.94&0.88&2.84&5.68\\ \hline
1&[0.6 0.001]&7&0.94&0.89&2.68&5.29&0.94&0.88&2.84&5.68\\ \hline
0.8&[0.6 0.001]&3&0.95&0.90&2.70&5.31&0.95&0.89&2.84&5.68\\ \hline
0.8&[0.6 0.001]&5&0.95&0.90&2.69&5.31&0.95&0.89&2.84&5.68\\ \hline
0.8&[0.6 0.001]&7&0.95&0.90&2.69&5.31&0.95&0.89&2.84&5.68\\ \hline
0.8&[.6 .3 0.001]\hspace{-4mm}&3&0.94&0.88&2.68&5.28&0.94&0.88&2.84&5.68\\ \hline
0.8&[.6 .3 0.001]\hspace{-4mm}&5&0.94&0.88&2.67&5.28&0.94&0.88&2.84&5.68\\ \hline
\end{tabular}
\end{center}
\caption{Comparision of theoretical and simulation values for SCNS}\label{netwok_validation}
\end{table}

\begin{table}[h]
\centering
\begin{tabular}{|l|l|l|l|l|}
\hline
\multicolumn{3}{|c|}{{ Configuration}} & \multirow{2}{*}{{ $P_{Busy}$}} & \multirow{2}{*}{{$ P_{Drop}$}} \\ \cline{1-3}
{ $P$} & { ${\mathcal{R}}$} & { $n$} & & \\ \hline
.6 & {[}0.5 0.0005{]} & 7 & 0.002755 & 0.03376 \\ \hline
.8 & {[}0.5 0.0005{]} & 7 & 0.000749 & 0.008509 \\ \hline
1 & {[}0.5 0.0005{]} & 7 & 0.00018 & 0.001907 \\ \hline
0.8&[0.6 0.001]&3&8.21E-05&8.44E-04\\ \hline
1&[0.6 0.001]&3&1.05E-05&1.00E-04\\ \hline
1.2&[0.6 0.001]&3&1.57E-06&1.42E-05\\ \hline
\end{tabular}
\caption{Effect on performance parameters with power variation}\label{power_change}

\end{table}
\subsection{SCNS - Optimal configuration}
We first study the performance of the system for different configurations.
 Table \ref{power_change} shows the busy probability $P_{Busy}$ and drop probability $P_{Drop}$ for different power profiles. It also considers different set of transmission rates. It can be noted that performance measures improve with increase in power. More importantly one can estimate the exact improvement of performance for a given increment in power. Table \ref{rate_change1} and \ref{rate_change2} present the sensitivity of performance measures with the change in the set of transmission rates. When an extra transmission rate is added to the existing set of transmission rates, there is a tremendous improvement in performance (see Table \ref{rate_change1}). On the other hand, performance of the system deteriorates with the decrease in maximum rate in transmission rate matrix (see Table \ref{rate_change2}) even after adding an extra rate. In all, one can use SCNS to derive optimal configuration.

 \begin{table}[h]
\centering
\begin{tabular}{|l|l|l|l|l|}
\hline
\multicolumn{3}{|c|}{{ Configuration}} & \multirow{2}{*}{{ $P_{Busy}$}} & \multirow{2}{*}{{ $P_{Drop}$}} \\ \cline{1-3}
{ $P$} & { ${\mathcal{R}}$} & { $n$} & & \\ \hline
0.8 & {[}0.6 0.001{]} & 3 & 8.21E-05 & 8.44E-04 \\ \hline
0.8 & {[}.6 .3 0.001{]} & 3 & 2.98E-06 & 2.73E-05 \\ \hline
0.8 & {[}0.6 0.001{]} & 5 & 8.54E-05 & 8.78E-04 \\ \hline
0.8 & {[}.6 .3 0.001{]} & 5 & 4.65E-06 & 4.25E-05 \\ \hline
\end{tabular}
\caption{Improvement in performance with an additional transmission rate}\label{rate_change1}
\centering
\end{table}

\begin{table}[h]\centering
\begin{tabular}{|l|l|l|l|l|}
\hline
\multicolumn{3}{|c|}{{ Configuration}} & \multirow{2}{*}{{ $P_{Busy}$}} & \multirow{2}{*}{{ $P_{Drop}$}} \\ \cline{1-3}
{ $P$} & { ${\mathcal{R}}$} & { $n$} & & \\ \hline
0.6 & {[}.5 0.0005{]} & 5 & 2.71E-03 & 3.32E-02 \\ \hline
0.6 & {[}.3 .1 0.0005{]} & 5 & 8.61E-03 & 1.14E-01 \\ \hline
0.8 & {[}.5 0.0005{]} & 5 & 7.28E-04 & 8.27E-03 \\ \hline
0.8 & {[}.3 .1 0.0005{]} & 5 & 4.41E-03 & 5.59E-02 \\ \hline
\end{tabular}
\caption{Impact on performance with an additional transmission rate
and decrease in maximum rate}\label{rate_change2}
\end{table}
\subsection{Testing of power law}

We now explore the efficiency and optimality of power law. In Section \ref{sec_opt_power_law}, it is theoretically established that affine power law is optimal for small cell network under certain assumptions. We demonstrate the supremacy of the power law, even in scenarios where the system does not satisfy the assumptions required by theory. 
\begin{table}[h]
\centering
\begin{tabular}{|l|l|l|l|l|l|l|l|l|}
\hline
\multicolumn{4}{|c|}{{\bf \begin{tabular}[c]{@{}c@{}}System\\ Configurations\end{tabular}}} & \multicolumn{2}{c|}{{\bf \begin{tabular}[c]{@{}c@{}}Without Power\\ Law\end{tabular}}} & \multicolumn{2}{c|}{{\bf \begin{tabular}[c]{@{}c@{}}With Power \\ Law\end{tabular}}} & \multicolumn{1}{c|}{\multirow{2}{*}{{\bf \begin{tabular}[c]{@{}c@{}}$P_{Drop}$ \\ $\hspace{-2mm}$\%change $\hspace{-3mm}$\end{tabular}}}} \\ \cline{1-8}
\multicolumn{1}{|c|}{$P$} & \multicolumn{1}{c|}{$\beta$} & \multicolumn{1}{c|}{$s_h$} & \multicolumn{1}{c|}{\begin{tabular}[c]{@{}c@{}}range \\$V$\end{tabular}} & \multicolumn{1}{c|}{$P_{Busy}$} & \multicolumn{1}{c|}{$P_{Drop}$} & \multicolumn{1}{c|}{$P_{Busy}$} & \multicolumn{1}{c|}{$P_{Drop}$} & \multicolumn{1}{c|}{} \\ \hline
0.85 & 2 & 0.4 & {[}20, 30{]} & 6.32E-04 & 7.21E-03 & 5.05E-04 & 5.66E-03 & 24.08 \\ \hline
1 & 2.5 & 0.4 & {[}20, 40{]} & 9.54E-03 & 1.57E-01 & 7.89E-03 & 1.22E-01 & 25.08 \\ \hline
1.2 & 2.5 & 0.4 & {[}20, 25{]} & 2.28E-04 & 2.98E-03 & 1.22E-04 & 1.50E-03 & 66.07 \\ \hline
1.6 & 3.1 & 0.5 & {[}20, 40{]} & 6.07E-04 & 8.47E-03 & 2.53E-04 & 3.22E-03 & 89.82 \\ \hline
1.6 & 2.1 & 0.5 & {[}20, 40{]} & 7.35E-03 & 1.20E-01 & 4.90E-03 & 7.36E-02 & 47.93 \\ \hline
\end{tabular}
\caption{Improvement in performance measure with optimal power law}\label{small_pbusy}

\end{table}

\begin{table}
\centering
\begin{tabular}{|l|l|l|l|l|l|l|l|l|}
\hline
\multicolumn{4}{|c|}{{\bf \begin{tabular}[c]{@{}c@{}}System\\ Configurations\end{tabular}}} & \multicolumn{2}{c|}{{\bf \begin{tabular}[c]{@{}c@{}}Without Power\\ Law\end{tabular}}} & \multicolumn{2}{c|}{{\bf \begin{tabular}[c]{@{}c@{}}With Power \\ Law\end{tabular}}} & \multicolumn{1}{c|}{\multirow{2}{*}{{\bf \begin{tabular}[c]{@{}c@{}}$P_{Drop}$ \\$\hspace{-3mm}$ \% change $\hspace{-3mm}$\end{tabular}}}} \\ \cline{1-8}
\multicolumn{1}{|c|}{$P$} & \multicolumn{1}{c|}{$\beta$} & \multicolumn{1}{c|}{$s_h$} & \multicolumn{1}{c|}{\begin{tabular}[c]{@{}c@{}}range \\$V$\end{tabular}} & \multicolumn{1}{c|}{$P_{Busy}$} & \multicolumn{1}{c|}{$P_{Drop}$} & \multicolumn{1}{c|}{$P_{Busy}$} & \multicolumn{1}{c|}{$P_{Drop}$} & \multicolumn{1}{c|}{} \\ \hline
1 & 2.5 & 0.6 & {[}20, 40{]} & 3.72E-02 & 6.71E-01 & 3.90E-02 & 6.62E-01 & -1.35 \\ \hline
1 & 3.1 & 0.5 & {[}20, 40{]} & 4.05E-02 & 7.24E-01 & 4.33E-02 & 7.40E-01 & -2.18 \\ \hline
\end{tabular}
\caption{System configuration (large $P_{Drop}$) when power law fails. }\label{large_Pbusy}

\end{table}

\begin{table}[h]
\centering
\begin{tabular}{|l|l|l|l|l|l|l|l|l|}
\hline
\multicolumn{4}{|c|}{{\bf \begin{tabular}[c]{@{}c@{}}System\\ Configurations\end{tabular}}} & \multicolumn{2}{c|}{{\bf \begin{tabular}[c]{@{}c@{}}Without Power\\ Law\end{tabular}}} & \multicolumn{2}{c|}{{\bf \begin{tabular}[c]{@{}c@{}}With Power \\ Law\end{tabular}}} & \multicolumn{1}{c|}{\multirow{2}{*}{{\bf \begin{tabular}[c]{@{}c@{}}$P_{Drop}$ \\\% change\end{tabular}}}} \\ \cline{1-8}
\multicolumn{1}{|c|}{$P$} & \multicolumn{1}{c|}{$\beta$} & \multicolumn{1}{c|}{$s_h$} & \multicolumn{1}{c|}{\begin{tabular}[c]{@{}c@{}}range \\ $V$\end{tabular}} & \multicolumn{1}{c|}{$P_{Busy}$} & \multicolumn{1}{c|}{$P_{Drop}$} & \multicolumn{1}{c|}{$P_{Busy}$} & \multicolumn{1}{c|}{$P_{Drop}$} & \multicolumn{1}{c|}{} \\ \hline
1 & 3 & 0.4 & {[}20, 40{]} & 2.47E-02 & 4.30E-01 & 2.37E-02 & 3.92E-01 & 9.24 \\ \hline
1.2&3&0.4&[20, 40]&8.25E-03&1.34E-01&6.67E-03&1.02E-01& 27.11\\ \hline
1.4&3&0.4&[20, 40]&3.26E-04&4.36E-03&1.75E-04&2.19E-03&66.25\\ \hline
\end{tabular}
\caption{Power law performing better for higher power}\label{Powerlaw_withpower}
\end{table}

Table \ref{small_pbusy} shows that power law works well for most of the cases especially when probability $P_{Drop},$ $P_{Busy}$ is small (at least of order $10^{-2}$). An improvement upto $89\%$ is noted for certain configuration (see Table \ref{small_pbusy}). Power law fails when $P_{Drop}$ takes large value in the order of $10^{-1}$ (see Table \ref{large_Pbusy}). The loss in performance due to power law is of orders less than 10\%. However, wireless networks generally don't operate in such scenarios due to low efficiency. Thus power law is useful in all `operating' case studies.

\begin{table}[h]
\centering
\begin{tabular}{|l|l|l|l|l|l|l|l|l|}
\hline
\multicolumn{4}{|c|}{{\bf \begin{tabular}[c]{@{}c@{}}System\\ Configurations\end{tabular}}} & \multicolumn{2}{c|}{{\bf \begin{tabular}[c]{@{}c@{}}Without Power\\ Law\end{tabular}}} & \multicolumn{2}{c|}{{\bf \begin{tabular}[c]{@{}c@{}}With Power \\ Law\end{tabular}}} & \multicolumn{1}{c|}{\multirow{2}{*}{{\bf \begin{tabular}[c]{@{}c@{}}$P_{Drop}$ \\\% change\end{tabular}}}} \\ \cline{1-8}
\multicolumn{1}{|c|}{P} & \multicolumn{1}{c|}{$\beta$} & \multicolumn{1}{c|}{$s_h$} & \multicolumn{1}{c|}{\begin{tabular}[c]{@{}c@{}}range \\$V$\end{tabular}} & \multicolumn{1}{c|}{$P_{Busy}$} & \multicolumn{1}{c|}{$P_{Drop}$} & \multicolumn{1}{c|}{$P_{Busy}$} & \multicolumn{1}{c|}{$P_{Drop}$} & \multicolumn{1}{c|}{} \\ \hline
1.6&2.1&0.5&[20, 40]&7.35E-03&1.20E-01&4.90E-03&7.36E-02& 47.93 \\ \hline
1.6 & 3.1 & 0.5 & {[}20, 40{]} & 6.07E-04 & 8.47E-03 & 2.53E-04 & 3.22E-03 & 89.82 \\ \hline
\end{tabular}
\caption{Improvement with Power Law for Smaller Values of $P_{Busy}$ }\label{Powerlaw_smallerPbusy}
\end{table}

\begin{figure}
\begin{minipage}{8cm}
\vspace{-5mm}
\includegraphics[width=8cm, height=12.5cm]{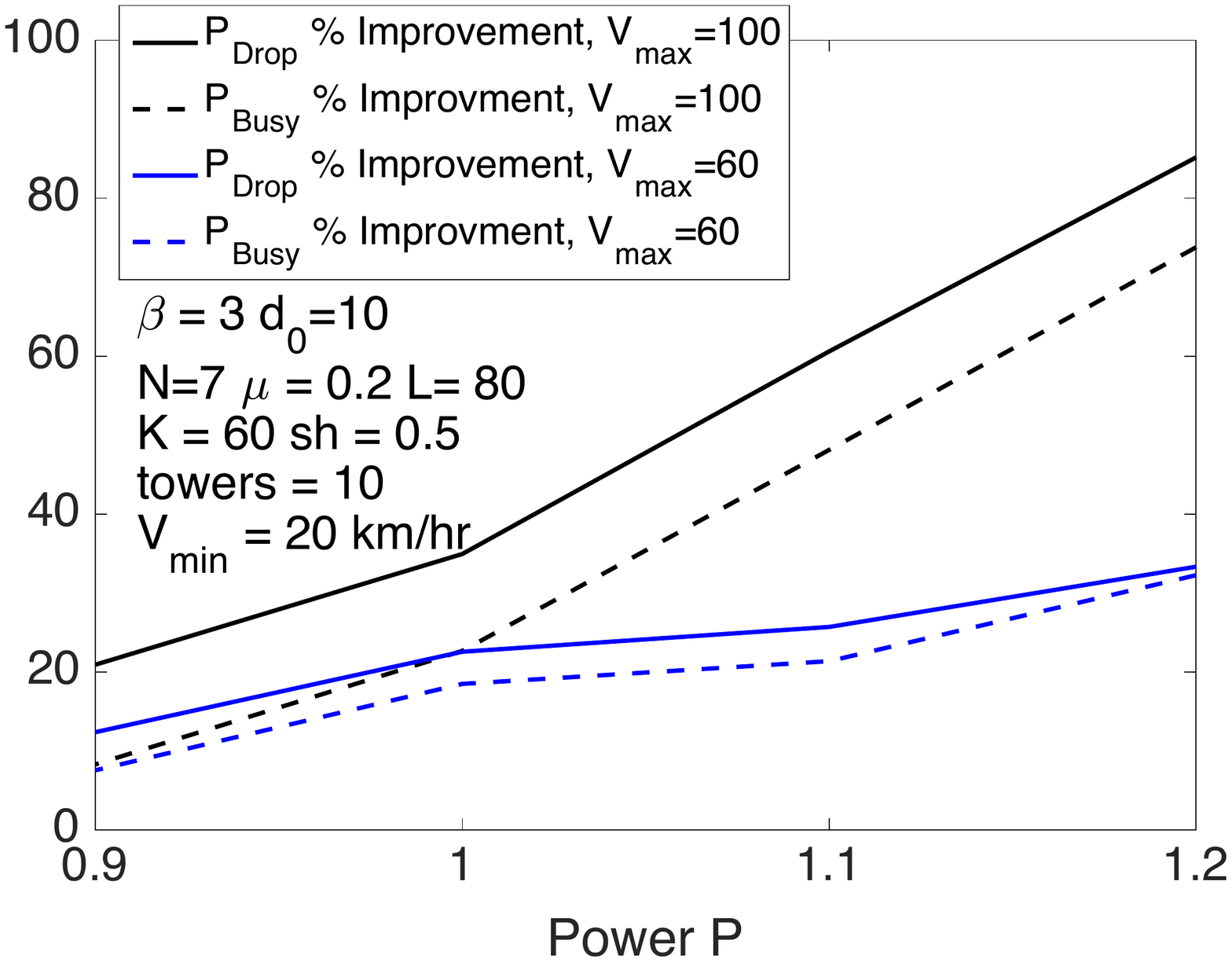}
\vspace{-38mm}
\caption{Normalized $\%$ improvement  for two cases \label{Fig_improve_and_Ps}}
\end{minipage}
\hspace{5mm}
\begin{minipage}{8cm}
\vspace{-2mm}
\includegraphics[width=8cm, height=11cm]{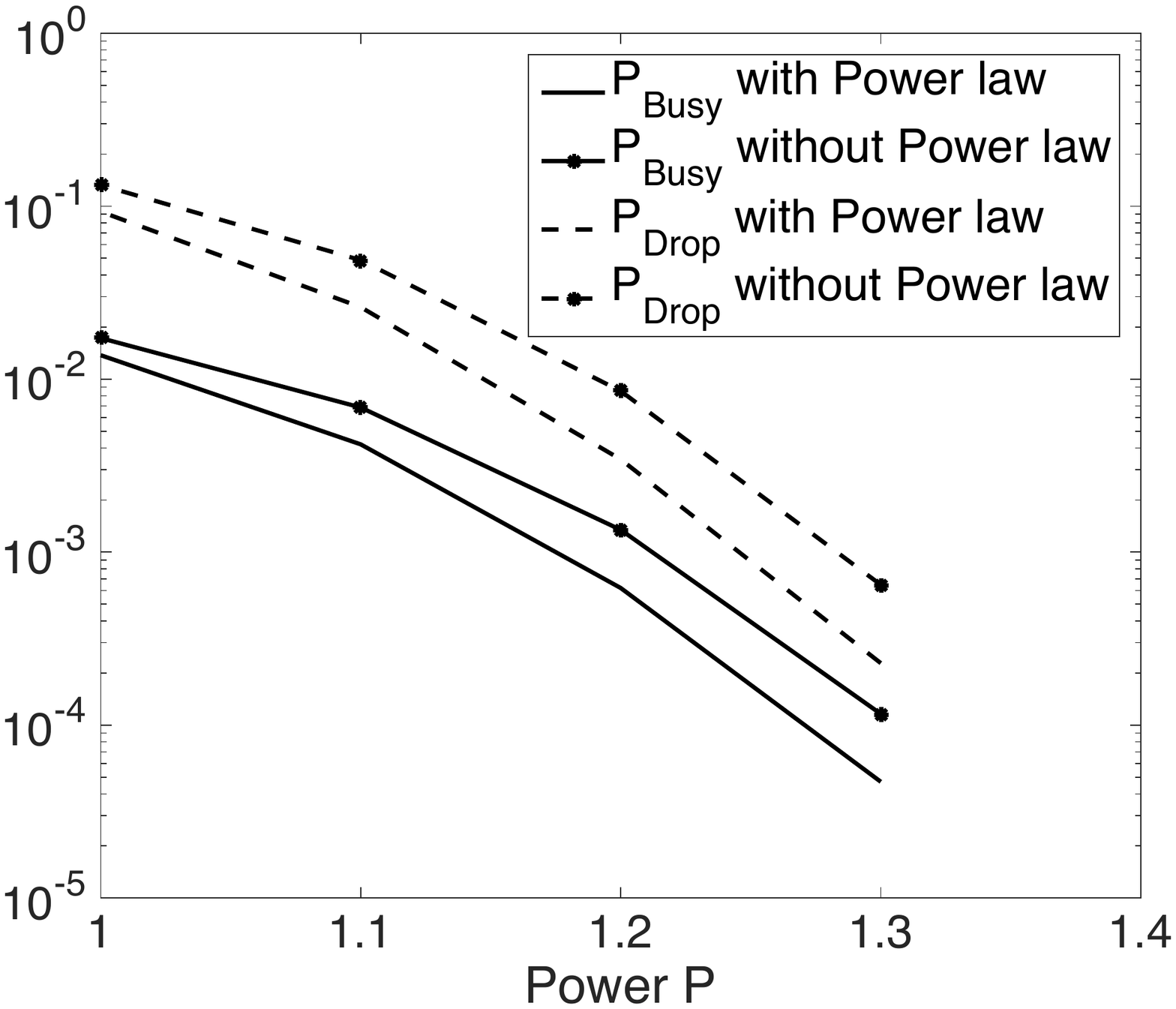}
\vspace{-32mm}
\caption{Block, drop probabilities with and without  linear power law,  when $V_{max = 100}$ \label{Fig_improve_and_Ps1}}
\end{minipage}

\end{figure}

Table \ref{Powerlaw_withpower} shows that the performance of power law improves with increase in power (upto 66\%). It is noted from the results of Table \ref{Powerlaw_smallerPbusy} that the improvement depends upon the path loss factor $\beta$. One has huge improvements for large path-loss scenarios.
Further, in all the test cases (irrespective of the configuration), we again notice a very good improvement when $P_{Drop}$ is small and is in the range of $10^{-2}-10^{-4}$ (see Tables \ref{Powerlaw_withpower}-\ref{Powerlaw_smallerPbusy}). We make a similar observation in the experiments of subsequent section.

We consider another example in Figures \ref{Fig_improve_and_Ps}-\ref{Fig_improve_and_Ps1}. We plot the normalized percentage improvement as a function of average power constraint $P$ in Figure \ref{Fig_improve_and_Ps}, while the block, drop probabilities are plotted in Figure \ref{Fig_improve_and_Ps1}. We consider two examples in  Figure \ref{Fig_improve_and_Ps}, the improvement is more when the range of user speeds is larger (black set of curves with $V_{max } = 100km/hr$).  From Figure \ref{Fig_improve_and_Ps1}, both drop probability and block probability improve significantly when power law is used (curves without markers are with power law).  From theory (Theorems \ref{Thrm_DiscretePowerLaw}-\ref{Thm_cts_law}) the disparity in the powers allocated to different speed classes, increases with increase in the range of user speed, and we observe from both the figures that this increase (in disparity) also increases the normalized improvement obtained using the power law.

\section{Numerical illustration: With a common rates set, ${\mathcal{R}}$ and interference}
\label{sec_onerateset}
In previous sections, we assumed that the (finite) set of transmission rate choices ${\mathcal{R}}$, depends upon the speed/transmit power of the user. We relax this assumption now and consider a big common set of transmission rate choices. System chooses a rate from the given set based on the received signal strength and rate set is the same for all users. One can {\it choose this common set of possible transmission rates, (${\mathcal R}$ and $N$), based on the practical channel coding schemes} that would be used in the network design.

We analyze this scenario using the simulator SCNS. We do not have theoretical support as of now for this scenario, however we note that the linear power law can improve the performance. We set power control rule to be:
\begin{eqnarray}
\label{Eqn_power_alpharule}
P(V ; \alpha) := \alpha {\bar P} + {\bar P} (1 - \alpha ) \frac{V}{E[V]} \mbox{ for any } \alpha \in [0, 1],
\end{eqnarray}and obtain the performance with best $\alpha$ using SCNS. If the set of transmission rates $\mathcal {R}$ is sufficiently big, we found that this rule provides a huge improvement in comparison with the scheme that allocates equal power (i.e., $\alpha = 1$) for all speeds.

The random trajectories of cars moving at random speeds, arriving at random positions with random service requirements is simulated in SCNS. For any given $\alpha$, the arriving user is allocated a transmit power according to rule (\ref{Eqn_power_alpharule}). As the car moves forward, its position changes and the received power changes accordingly. At any point the best rate among the rates of ${\mathcal{R}}$, which is below received SNR (using low SNR approximation (\ref{Eqn_cap}) for capacity) is chosen as the transmission rate.
\begin{table}[h]
\centering
\begin{tabular}{|l|l|l|l|l|l|l|l|l|}
\hline
\multicolumn{3}{|c|}{{\bf \begin{tabular}[c]{@{}c@{}}System\\ Configurations\end{tabular}}} & \multicolumn{2}{c|}{{\bf \begin{tabular}[c]{@{}c@{}}Without Power\\ Law ($\alpha = 1$) \end{tabular}}} & \multicolumn{2}{c|}{{\bf \begin{tabular}[c]{@{}c@{}}With Power \\ Law ($\alpha = .7$)\end{tabular}}} & \multicolumn{1}{c|}{\multirow{2}{*}{{\bf \begin{tabular}[c]{@{}c@{}} $\hspace{-3mm}$ \% Impr \\ \\ (in $P_{Drop}$) $\hspace{-3mm}$ \\ \end{tabular}}}} \\ \cline{1-8}
\multicolumn{1}{|c|}{$\bar{P}$} & \multicolumn{1}{c|}{$\beta$} & \multicolumn{1}{c|}{\begin{tabular}[c]{@{}c@{}}range \\$V$\end{tabular}} & \multicolumn{1}{c|}{$P_{Busy}$} & \multicolumn{1}{c|}{$P_{Drop}$} & \multicolumn{1}{c|}{$P_{Busy}$} & \multicolumn{1}{c|}{$P_{Drop}$} & \multicolumn{1}{c|}{} \\ \hline
0.75 & 3.5 & {[}20, 100{]} & 6.36E-05 & 4.40E-04 & 3.92E-05 & 2.54E-04 & 53 \\ \hline
0.75 & 3.0 & {[}20, 100{]} & 4.27E-04 & 3.20E-03 & 3.06E-04 & 2.15E-03 & 39 \\ \hline
0.75 & 2.5 & {[}20, 100{]} & 2.79E-06 & 1.72E-05 & 1.53E-06 & 8.91E-06 & 62 \\ \hline \hline
0.7 & 35 & {[}20, 100{]} & 1.77E-03 & 1.43E-02 & 1.33E-03 & 1.00E-02 & 35 \\ \hline
0.7 & 3 & {[}20, 100{]} & 3.93E-04 & 2.93E-03 & 3.16E-04 & 2.21E-03 & 29 \\ \hline
0.7 & 2.5 & {[}20, 100{]} & 2.98E-05 & 1.99E-04 & 1.81E-05 & 1.14E-04 & 54 \\ \hline
\hline \hline
0.7 & 3.5 & {[}5, 100{]} & 1.02E-03 & 7.30E-03 & 7.21E-04 & 4.62E-03 & 45 \\ \hline
0.7 & 3 & {[}5, 100{]} & 2.03E-04 & 1.34E-03 & 1.37E-04 & 8.16E-04 & 48 \\ \hline
0.7 & 2.5 & {[}5, 100{]} & 1.28E-05 & 7.55E-05
 & 8.68E-06 & 4.66E-05 & 47 \\ \hline
\end{tabular}
\caption{ Improvement due to Power Law with common rates set
${\mathcal{R}} = \{0.8:-0.035:0.03 , 0.011:-0.004:0.003 \} $ and $|{\mathcal{R}} |= 26$
Other system parameters: $s_h = 0.4$, $n = 10$,
$\delta = 0.04$, $L = 70$, $V_{min}=20kmph$, $V_{max} = 100kmph$. }\label{common_rates_set}
\end{table}

We tabulate the improvement in performance with affine power law (\ref{Eqn_power_alpharule}) and a good $\alpha$ for many configurations in Table \ref{common_rates_set}.
We notice a significant improvement in performance for all cases. However, this improvement requires the availability of 26 different rate choices as mentioned in the caption of the table. This requirement is mandatory, because using this large set, we can create the effect of significantly different (virtual) cell sizes for different speed classes. In next subsection, we consider an example with common rates set in presence of interference. We notice significant improvement in this case also (see Table \ref{table_interference}).
 Thus by using a big set of transmission rates, we are able to illustrate a very good improvement in performance.
{\it It is practically impossible to create different (physical) cell sizes for different speed classes, however one can probably work with a large set of transmission rates. And hence this can be a good design alternative. }

\begin{table}
\centering
\resizebox{18.cm}{!}{
\begin{tabular}{|l|l|l|l|l|l|l|l|l|}
\hline
 { \bf System Config. }
 & \multicolumn{2}{c|}{{\bf \begin{tabular}[c]{@{}c@{}}Without Power Law \\ $\alpha = 1$ \end{tabular}}}
 & \multicolumn{2}{c|}{{\bf \begin{tabular}[c]{@{}c@{}}With Power Law \\ $\alpha < 1$ as in System Config. column \end{tabular}}}
 & {\bf \begin{tabular}[c]{@{}c@{}} \% Impr \\ (in $P_{Drop}$) \\ \end{tabular} } \\ \cline{1-6} \hline
 & & & && \\
\begin{tabular}[c]{@{}c@{}} $L$ \, \,\, $\alpha$ \, \, $\beta$ \, Rates \end{tabular} &\begin{tabular}[c]{@{}c@{}} Single Cell \\$P_{Busy}$(Interference) \end{tabular} & \begin{tabular}[c]{@{}c@{}} System \\$P_{Drop}$(Interference) \end{tabular}
 & \begin{tabular}[c]{@{}c@{}} Single Cell \\$P_{Busy}$(Interference) \end{tabular} & \begin{tabular}[c]{@{}c@{}} System \\$P_{Drop}$(Interference) \end{tabular}
 & \\

 & &&&& \\ \hline

 $ 80 $ \, $ 0.6$ \, $ 2.5$ \, ${\mathcal R}_1$ &7.36 (7.73, 8.99) E-04 & 5.77 (6.08, 7.13) E-03 & 5.36 (5.66, 8.52) E-04 & 4.00 (4.25, 6.52) E-03 & 36 (35, 09) \\ \hline


 $ 80 $ \, $ 0.6$ \, $ 2.5$ \, ${\mathcal R}_2$ &4.89 (5.17, 5.99) E-04 & 3.76 (3.99, 4.67) E-03 & 3.11 (3.33, 4.96) E-04 & 2.26 (2.44, 3.70) E-03 & 50 (48, 24) \\ \hline \hline \hline

$ 70 $ \, $ 0.7$ \, $ 2.5$ \, ${\mathcal R}_1$ &6.66 (7.09, 8.90) E-05 & 5.29 (5.64, 7.20) E-04 & 3.80 (4.60, 8.65) E-05 & 2.88 (3.51, 6.80) E-04 & 59 (47, 06) \\ \hline

$ 70 $ \, $ 0.7$ \, $ 2.5$ \, ${\mathcal R}_2$ &3.81 (4.00, 5.37) E-05 & 2.96 (3.11, 4.22) E-04 & 1.82 (2.00, 4.62) E-05 & 1.34 (1.49, 3.53) E-04 & 75 (70, 18) \\ \hline \hline \hline

$ 70 $ \, $ 0.7$ \, $ 3.0$ \, ${\mathcal R}_2$ &6.09 (6.11, 6.35) E-04 & 5.34 (5.36, 5.59) E-03 & 3.37 (3.36, 3.92) E-04 & 2.80 (2.78, 3.29) E-03 & 62 (63, 52) \\ \hline
\end{tabular}}
\caption{\label{table_interference} With interference and `non-uniform' speeds: $\delta = 0.01$, $s_h = 0.4$, $\mu=0.2$, $K=60$,
$\lambda = 0.7n$, $d_0 = 10$, ${\bar P} = 0.7$ and $\# \mbox{ of towers }n = 10$ and noise variance $\sigma^2 = 0.5$ or $ 0.1$. 
In each results column, the quantities inside the round brackets are the performance measures with interference, while the one outside is without interference, all for the same configuration.
 For example $\%$ improvement is 50 for the configuration of  second row, 
without considering interference while it equals 48 and 24 respectively with noise variances $\sigma^2 = 0.5$ and $ 0.1$.
 }
\end{table}

\subsection{With Interference and `non-uniform' speeds} In this subsection, we further consider the effects of interference. Our theoretical results are valid for any distribution of velocity/speed. We also consider speeds that are not uniformly distributed. We assume that the velocity is a conditional Gaussian random variable as below:
$$
V = \left . {\cal N} \left ( \frac{V_{min}+V_{max}}{2}, 10 \right ) \right | {\cal N} \in [V_{min}, \, V_{max}] ,
$$where ${\cal N} (m, v) $ is a Gaussian random variable with mean $m$ and variance $v$. Basically we generate Gaussian random variables and consider only the values that are within $[V_{min}, \, V_{max}] .$ We set $V_{min} = 40kmph$, $V_{max} = 100kmph$ for this set of experiments.
We assume that the towers of the cells other than the serving cell cause interference to the user. The total interference term, ${\cal I}$, is computed using the
received powers at the user under consideration, received from all the towers that serve the other users currently utilizing the system. These received powers are attenuated values of the power allocated according to rule (\ref{Eqn_power_alpharule}), after distance based attenuation as according to equation (\ref{Eqn_rates}). Hence the interference term and the SINR (Signal to Noise and Interference ratio) of the tagged user equals:
\begin{eqnarray}
{\cal I} &=& \sum_{\begin{array}{llll}&\mbox{users of other cells} \\ &\mbox{ in same channel} \end{array} } P(V_i, \alpha) (d_i / d_0)^{-\beta}, \nonumber \\
SINR &=& \frac{P (V_{tagged}, \alpha) (d_{tagged }/ d_0)^{-\beta} }{1 +\frac{ {\cal I}}{\sigma^2}}, \label{Eqn_interference_SINR}
\end{eqnarray}where $V_i$ is the velocity of user $i$, whose tower is at distance $d_i$ from the tagged user and where $\sigma^2$ is the thermal noise variance.
Recall, by the convention of our paper, that ${\bar P}$ actually represents the transmit power divided by the thermal noise variance.
Here, we are estimating the SINR in place of SNR and the best rate (in ${\mathcal{R}}$) below the SINR is chosen for transmission, similar to previous section.

We consider two different case studies and the respective set of (common) rate choices are (respectively of sizes 30 and 45):
\begin{eqnarray*}{\mathcal{R}}_1 &=& \{0.83:-0.03:0.01, \, 0.007, \, 0.003 \} , \\ {\mathcal{R}}_2 &=& \{0.80:-0.02:0.02 , \, 0.009:-0.002: 0.001 \}. \end{eqnarray*}	
Basically when the path-loss factor is large one requires more number of rate choices.

The results are tabulated in Table \ref{table_interference}.
Every sub-column (other than system-configuration) has three entries. {\it The quantities inside the round brackets are the performance measures with interference, while the one outside is without interference, all for the same configuration.}
 The two entries within round brackets are the results with noise variance 0.5 and 0.1 respectively. As noted from the table, the power law improves the performance even for non-uniform speeds.
The performance of the system degrades with interference, however a significant improvement is noted with power law (entries inside the round brackets of the last column).
We also observe that improvement increases with a bigger set of rates (rows 2, 4 and 5 (with ${\mathcal R}_2$) in Table \ref{table_interference}). Good $\alpha$ turns out to be smaller for bigger cell sizes (rows 1 and 2). This indicates a bigger disparity in powers allocated to various speeds.

We considered many more test cases and found that the power law provides good improvement. We noticed that the percentage  improvement increases as the SINR increases or equivalently as $P_{Drop}$ decreases.  

The percentage improvement is reduced when one considers the effects of interference. Nevertheless the improvement has not diminished completely,   but rather is sufficiently significant even with interference. 
The simulations show an improvement up to 52\% (and in many
cases well above 9\%) even in the presence of (time varying and random)  strong interference (columns of Table \ref{table_interference} with $\sigma^2 = 0.1$).
In presence of smaller interference (with $\sigma^2 = 0.5$) the percentage improvement is much higher (well above 35 \% in all cases). 
 Further, the impact of considering interference on reduction in percentage improvement is higher with smaller path losses (first four rows with $\beta = 2.5$ in Table \ref{table_interference}).

\section{Conclusions}\label{sec_conc}
Mobility management is a challenging issue in HetNet deployments, where the main objective is to reduce call drops and network signalling flow, optimize traffic scheduling, and achieve resource optimization. As an example, a common and currently adapted solution for high speed trains is to densify the network along the railway tracks to combat the large penetration loss (coverage holes). However, this will increase HO frequency due to smaller site-to-site distance. Another way is to increase the transmission power of the BSs, which also helps in reducing the large penetration losses. However, this results in high energy consumption (and more interference) and neither of these solutions are cost-effective.
Previously in literature, for efficient design, it was suggested to vary inter BS distances based on the speed of the user. However, it is practically infeasible to do so. Alternately, we have proposed in this paper to reflect the speed based (optimal) design variations in the allocated powers.

We have obtained a closed form expression for optimal power control (optimal for load factor and/or busy probability), which allocates different transmit powers to different user (speed) classes, for any given average power constraint and the cell size. The optimal control ensures larger power to higher speeds and the differences in the powers allocated increase with path loss factor and the disparities in the speeds.
It is noteworthy that whenever an accurate user speed estimate is made available, one can obtain a very fine optimal power control with speed. We have shown that the optimal power varies linearly with the user speed and this result is obtained using Hamilton-Jacobi-Bellman equations.

We have also observed via numerical simulations that, there is large improvement in busy probability and load factor when optimal power control is used instead of equal powers for all users. The improvement in performance increases as the number of user classes increases and one obtains the best improvement with continuous optimal power law. The systems with large (path) losses and or the ones which support wide variations of user speeds,
improve significantly with optimal power control.

Further, a system level simulator has been built on which exhaustive simulations were conducted.
We have derived the theory under certain assumptions, however the performance of linear power law is estimated using system level simulator for more general set-up.
We have observed that power law outperforms the counter part of the systems that uses equal power for all speeds, in all the scenarios with moderate to high SNR conditions.
For the system with interference, the power law out-performs by significant margin (even up to 70\%) while for systems without interference we noticed an improvement even up to 89\% percentage. Achieving these results offers hope that such a smart mechanism can be designed around small cell integration in HetNets. This integration can further help the notion of CSB proposed by LTE standards.

We also establish that, for larger cells to work more efficiently, one has to boost the total transmit power
used in the system. We showed that this boosting depends upon the path loss factor beta and called it as beta+ scaling.
We finally propose a joint cost to obtain optimal cell size, when the network employs optimal power allocation as well as beta+ scaling.

\appendix

\section{ Maximum velocity supported by system}\label{powerassumption}
In this section, we calculate a limit on the speeds that can be supported, by a system with maximum $N$ possible transmission rates and with transmission power $P$. This is equivalent of \cite[Theorem 2]{Perf} obtained for systems with `maximal' transmission rates.
User entering at $-L$ when moving with speed $V$, can transfer in a cell, at maximum (see the discussions while deriving $\psi_n$ given by (\ref{Eqn_psin}) and using (\ref{Eqn_rates}))
{\small
\begin{eqnarray}
\label{Eqn_gdL}
 g^N(L) := \frac{ L}{N V} \sum_{m = -N}^{N} r_m = C_{h,ho} \frac{P L^{1-\beta}}{\mu V} ,
\end{eqnarray}}bytes of information, out of which $s_h$ are used for HO purposes. So, useful communication is possible only when
 $ g^N(L) > s_h$ with probability one.
With $\beta > 1$ (the practical range of path loss factors), $g^N(L)$ reduces with $L$ and so useful communication is not possible for any cell size if
$ g^N (Nd_0) $ itself is less than $s_h$ and hence we have:
\begin{thm}
\label{Thrm_one}
When $\beta > 1$, there exists a limit on the maximum velocity that can be supported by the system for a given power $P$ {
$$
V_{lim} (P) := \frac{1}{ s_h} r_0 \sum_{n = -N}^N \phi_n^{-\beta} P d_0^{1-\beta}N^\beta
 . \ \ \Box
$$}
\end{thm}

From the above theorem, given a set of system parameters, useful communication can be achieved by increasing the power $P$. It is assumed while deriving optimal power law that transmission power $P$ is large enough to ensure useful communication.

\section{Proof of Theorem}
{\bf Proof of Theorem \ref{Thrm_DiscretePowerLaw}:}
The average power constraint can be satisfied with equality by substituting
$P_I = \frac{1}{p_I} \left ({\bar P} - \sum_{i < I} p_i P_i \right ),$ and {then (with ${\bf P}_{-I} := (P_1, \cdots, P_{I-1})$)

{\small \vspace{-6mm}
\begin{eqnarray*}
\rho(L; {\bar P}) 	&:=& \rho \left (L, \left ({\bf P}_{-I}, \frac{ {\bar P} - \sum_{i < I} p_i P_i }{p_I} \right ) \right ) \\
& = &
 \frac{\lambda L^2}{K} \sum_{i < I} p_i \Upsilon_i \left ( C_{b,e} + C_{b,h} \frac{1 - \delta_i C_{e,ho}}
{ \delta_i C_{h,ho} - \mu s_h} \right ) + \frac{\lambda L^2}{K} p_I \Upsilon_I
 \left (C_{b,e} + C_{b,h} \frac{1 - {\bar \delta} ({\bf P}_{-I}) C_{e,ho}}{ {\bar \delta} ({\bf P}_{-I}) C_{h,ho} - \mu s_h} \right ), \\
{\bar \delta} ({\bf P}_{-I}) &:=& \frac{1}{p_I}\left ( {\bar P} - \sum_{i < I}p_i P_i \right ) \Upsilon_I L^{1-\beta}.
\end{eqnarray*} }
We obtain the optimizer of the load factor $\rho$ via the zeros of the derivatives (if they exist).
The partial derivatives (for all $i < I$) are given by: 
%

 {\small \vspace{-5mm}\begin{eqnarray}
 \label{Eqn_partialDerivative}
 \frac{d \rho}{d P_i} & =& \frac{\lambda}{K} C_{b,h} p_i \Upsilon_i^2 L^{3-\beta}
 \frac{ C_{e,ho} \mu s_h - C_{h,ho}}
 {\left ( \delta_i C_{h,ho} - \mu s_h \right )^2}
- \frac{\lambda}{K} C_{b,h} p_i \Upsilon_I^2 L^{3-\beta}
 \frac{ C_{e,ho} \mu s_h - C_{h,ho} }
 {\left ( {\bar \delta} ({\bf P}_{-I}) C_{h,ho} - \mu s_h \right )^2} \nonumber \\
 & = & \frac{\lambda}{K} C_{b,h} p_i L^{3-\beta} \left (C_{e,ho} \mu s_h - C_{h,ho} \right ) \left ( \frac{ \Upsilon_i^2 }
 {\left ( \delta_i C_{h,ho} - \mu s_h \right )^2} - \frac{ \Upsilon_I^2}
 {\left ({\bar \delta} ({\bf P}_{-I}) C_{h,ho} - \mu s_h \right )^2} \right ).
\end{eqnarray}}

One can easily obtain the zero of equation (\ref{Eqn_partialDerivative}), i.e., the equilibrium point.
Equating the partial derivatives to zero, $\partial \rho / \partial P_i = 0$ for all $i$, the optimal power control ${\bf P} (L, {\bar P})$, for a given cell size and average power constraint ${\bar P}$ equals:

{\small \vspace{-4mm} \begin{eqnarray}
\label{Eqn_OptPowerRaw}
{\bf P}^* (L) \hspace{-2mm}&=& \hspace{-2mm}
\left [ \hspace{-2mm}
\begin{array}{llllllll}
-1 &0 &0 &\cdots &0 & 1 \\
0 & -1 &0 & \cdots &0 & 1 \\
&& \vdots \\
0 &0 &0 &\cdots &-1 & 1\\
p_1 & p_2 &p_3 &\cdots &p_{I-1} & p_I
\end{array}\hspace{-2mm} \right ]^{-1} \hspace{-4mm}\frac{\mu s_h L^{\beta-1}}{\Upsilon_I C_{h,ho}} {\bf v}_\mu \nonumber\hspace{6mm} \\
{\bf v}_\mu &:=& \left [ \begin{array}{llllll} \frac{ \Upsilon_1 - \Upsilon_I }{\Upsilon_1 }, &
 \frac{\Upsilon_2 - \Upsilon_I }{\Upsilon_2}, & \cdots,
 &
\frac{\Upsilon_{I-1} - \Upsilon_I }{\Upsilon_{I-1} }, &
 {\bar P} \end{array} \right ]. \hspace{5mm}
\end{eqnarray} }
This can be solved easily and }{This problem can be solved if the power constraint ${\bar P}$ is
sufficiently large (see {\cite{LongTR}}).
The optimal power control ${\bf P}^* (L; {\bar P})$, for a given cell size and power constraint ${\bar P}$ equals (for all $i$):}

{\small \vspace{-5mm}\begin{eqnarray*}
P_i^* (L; {\bar P}) &=& {\bar P} + \frac{\mu s_hL^{\beta-1}}{ C_{h,ho}} \sum_{j < I} p_j \left ( \frac{1}{\Upsilon_I} - \frac{1}{\Upsilon_j} \right ) - \frac{\mu s_h L^{\beta-1}}{ C_{h,ho}}\left ( \frac{1}{\Upsilon_I} - \frac{1}{\Upsilon_i} \right ).
\end{eqnarray*}}
This simplifies to (\ref{Eqn_OptPower}). Differentiating (\ref{Eqn_partialDerivative}) again, the Hessian matrix equals:
\begin{eqnarray*}
 \frac{ 2 C_{h,ho} \lambda C_{b,h} L^{4-2\beta} \left ( C_{h,ho} - C_{e,ho} \mu s_h \right ) \Upsilon_I^3}
 {K \left ({\bar \delta} ({\bf P}_{-I}) C_{h,ho} - \mu s_h \right )^3} {\cal P}_V,
\end{eqnarray*} where ${\cal P}_V$ is defined in the hypothesis of the theorem.
This is a positive definite matrix under the given assumptions and hence ${\bf P}^*$ given by (\ref{Eqn_OptPower}) is indeed a minimizer.
$\Box$

\section{ Expected overall service time $B$}\label{sec_expectedB}
Let $Z = (H_O, V)$ represent a joint random variable in which the first component is an indicator that it is due to HO, i.e., $H_O = 1$ implies it is an HO call.
The following are the events given that a call has arrived. We first consider HO case. Let $\prvV$ and $\prvHO$ represent
the velocity and the call type (HO or external) in the previous cell before the Handover to the current cell. That is, $\prvHO = 1$ implies it was a HO call in the previous cell and is again handed over to the current cell. For any set ${\cal A}$, since
the velocity of user remains constant during the call:
{\small \begin{eqnarray}
P (H_O = 1, V \in {\cal A}) = P(H_O = 1, \prvV \in {\cal A} )= P (H_O = 1, \prvV \in {\cal A}, \prvHO = 0) + P (H_O = 1, \prvV \in A, \prvHO = 1) \nonumber \\ \label{Eqn_onne}
\end{eqnarray}}
By stationarity $P (\prvHO = 0) = P (H_O = 0)$ and this equals the probability that a call arrived is
a new call. By ergodicity this equals $\lambda_L / {\bar \lambda}.$
Thus, by conditioning first on $\prvHO$ and then on $\prvV$ we obtain:
{\small \begin{eqnarray}
P (H_O = 1, \prvV \in {\cal A}, \prvHO = 0)
 &=& P (H_O = 1, \ \prvV \in {\cal A} | \prvHO = 0) P (\prvHO = 0) \nonumber \\
 &= & \int_{A} P_{e,ho} (v) p_V (v) dv \, \frac{\lambda_L}{{\bar \lambda}}. \label{Eqn_two}
\end{eqnarray}} Considering infinitesimal interval $vdv := [v-dv, v+dv]$ and conditioning as in the previous equation we obtain:
{\small\begin{eqnarray}
 P(H_O = 1, \prvV \in vdv, \prvHO = 1) &=& P(H_O = 1 | \prvV \in vdv, \prvHO = 1) P(\prvHO = 1, \prvV \in vdv) \nonumber \\ &\approx &
P_{h,ho} (v) P(\prvHO = 1, \prvV \in vdv). \label{Eqn_three}
\end{eqnarray}}
By stationarity (and $S$ being memoryless), $P(\prvHO = 1, \prvV \in vdv) = P(H_O = 1, V \in vdv)$ and thus by substituting (\ref{Eqn_two}), (\ref{Eqn_three}) into (\ref{Eqn_onne}),
\begin{eqnarray}
P (H_O = 1, V \in vdv)\approx \frac{ P_{e,ho} (v) p_V (v) dv } {1 - P_{h,ho}} \frac{\lambda_L}{{\bar \lambda}}.
\label{Eqn_one}
\end{eqnarray}
 In a similar
way,
 \begin{eqnarray}
 P(H_O = 0, V \in vdv) & =& P (V \in vdv | H_O = 0) \frac{\lambda_L}{{\bar \lambda}} \nonumber \\ &\approx& p_V (v) dv \frac{\lambda_L}{{\bar \lambda}}. \label{Eqn_zero}
 \end{eqnarray}
 Note that service requirements $S$ is independent of $ Z= (H_O, V)$. Hence,
 using equation (\ref{Eqn_Ctsbarlambda}) and equations (\ref{Eqn_onne})-(\ref{Eqn_zero})
 and conditioning on $Z$, the average of the overall service time equals the equation (\ref{Eqn_Ctsbarb}).
\end{document}